# Emulating an Integrated Assessment Model to Project Long-Term Emissions and Carbon Dioxide Removal Pathways to 2300

Katsumasa Tanaka[1,2,*], Weiwei Xiong[1,*], Leon Merfort[3], Nico Bauer[3]

[1] Laboratoire des Sciences du Climat et de l'Environnement (LSCE), IPSL, CEA-CNRS-UVSQ, Université Paris-Saclay, Gif-sur-Yvette, France

[2] Earth System Division, National Institute for Environmental Studies (NIES), Tsukuba, Japan

[3] Potsdam Institute for Climate Impact Research (PIK), Potsdam, Germany

* These authors contributed equally to this work.

Corresponding authors: katsumasa.tanaka@lsce.ipsl.fr and weiwei.xiong@lsce.ipsl.fr

## Disclaimer

This working paper, prepared for the 30th Annual Conference of the European Association of Environmental and Resource Economists (EAERE) (16-19 June, 2025, Bergen, Norway), presents preliminary results from an ongoing research focusing on the development and application of IAM emulators for generating long-term climate scenarios. The results and analyses presented here are subject to further refinement and should not be considered final. We welcome comments and suggestions that can inform the next stages of this research.

## Abstract

Integrated Assessment Models (IAMs) are essential tools for exploring climate change mitigation pathways, but they typically generate emission scenarios only until 2100. To address the growing need for longer-term scenarios beyond 2100, we developed an emulator of the IAM REMIND-MAgPIE. This emulator, emIAM, captures the behavior of the IAM by using marginal abatement cost (MAC) curves for three major greenhouse gases (GHGs), two sectors, and five carbon dioxide removal (CDR) techniques. We coupled emIAM with the reduced-complexity climate model ACC2 to generate a suite of internally consistent scenarios extending to 2300, targeting a range of temperature stabilization levels with varying overshoot profiles. These scenarios go beyond previously employed idealized approaches by incorporating the emulated dynamic response of the IAM and by explicitly characterizing CDR deployment pathways. Our scenarios serve as input to Earth System Models (ESMs) investigating the long-term consequences of climate change mitigation strategies, particularly the implications of CDR deployment and associated Earth system dynamics over centennial timescales. Furthermore, the modular design of the IAM emulator facilitates coupling with other reduced-complexity climate models, broadening its applicability for exploring climate-economic interactions.

## 1. Introduction

CDR techniques are no longer considered optional but are now recognized as an essential component of pathways towards achieving net-zero $CO_2$ and GHG emission targets (IPCC 2023). With the 1.5 °C warming threshold likely to be crossed soon (Schmidt 2024; Bossy et al. 2024; Goessling et al. 2024), CDR techniques are becoming increasingly relevant for potentially reversing, or at least mitigating, warming if efforts continue to pursue the 1.5 °C warming target with overshoot (Schleussner et al. 2024). While a diverse portfolio of CDR techniques has emerged (Smith et al. 2024), afforestation remains the most established, currently representing the vast majority of deployed CDR. Beyond afforestation, other techniques such as bioenergy with carbon capture and storage (BECCS), direct air capture and storage (DACCS), industry CCS, and ocean alkalinity enhancement (OAE) are gaining increasing attention. IAMs often rely heavily on large-scale deployment of BECCS in scenarios with limited early mitigation followed by a temperature overshoot (Riahi et al. 2021). However, the heavy reliance on this technique raises concerns about potential land-use conflicts with food production and biodiversity conservation (Smith et al. 2016; Gambhir et al. 2019). DACCS is another technically viable, yet currently cost-prohibitive, option of CDR. More recently, OAE is gaining attention, leveraging specific minerals to enhance ocean $CO_2$ uptake by altering ocean alkalinity.

Despite their prominence in IAMs, the long-term climate effects of these CDR techniques have been subject to limited investigation within ESMs (Melnikova et al. 2023b), leaving their physical and biogeochemical impacts largely unexplored over centennial timescales. The typical IAM time horizon ending in 2100 (O'Neill et al. 2016) hinders such long-term exploration (Easterling et al. 2024). To address this gap, we developed an emulator of an IAM and coupled it with a reduced-complexity climate model. This coupled model allows us to generate a suite of novel emission scenarios extending to 2300, including detailed CDR deployment pathways. Our scenarios cover a range of long-term temperature target levels (2 °C, 1.5 °C, and 1 °C) with differing degrees of temperature overshoot (up to about 2 °C above the target level), representing various CDR use cases over multiple centuries. We focus on global levels by assuming that the emission or forcing spatial patterns are fixed from 2100 onwards, while acknowledging their potential further evolutions and associated uncertainties. These scenarios will allow ESMs to examine the long-term consequences of different CDR deployment strategies, including their effectiveness in achieving various temperature targets and the potential impacts on physical and biogeochemical processes in the Earth system. Such ESM simulations further enable climate impact assessments in the long term. Through the use of emIAM, our extended scenarios until 2300 are more consistent with and more strongly linked to the scenarios until 2100 directly generated by IAMs than previously employed idealized long-term scenarios that rely primarily on linearly declining emission trajectories (Meinshausen et al. 2020).

The remainder of this paper is structured as follows. Section 2 introduces a brief description of REMIND-MAgPIE, the IAM primarily considered in this study, and details the methodology for developing emIAM, including the underlying MAC curves and other key properties. This section also describes the reduced-complexity climate model ACC2 to be coupled with emIAM and presents the evaluations of the performance of the emulator. Section 3 presents the suite of extended scenarios till 2300, showcasing the detailed profiles of GHG emissions and CDR deployment pathways under a range of temperature targets and overshoot profiles. Section 4 discusses the broader implications of these long-term scenarios, explores the potential future applications of emIAM, and concludes the paper.

## 2. Methods

### 2.1. REMIND-MAgPIE

This study employs REMIND-MAgPIE (Kriegler et al. 2017; Bauer et al. 2020a, b; Merfort et al. 2023), a global Integrated Assessment Model (IAM) developed at the Potsdam Institute for Climate Impact Research (PIK) in Germany, which represents the complex interplay between the economy, energy system, land use, and climate change. REMIND-MAgPIE simulates the welfare-maximizing evolution of the global socioeconomic system disaggregated into twelve world regions, with a particular focus on the energy and land-use sectors under specified constraints related to population growth, technological change, policy implementation, and climate target. REMIND consists of a macroeconomic growth model coupled with a bottom-up, engineering-based energy system model. REMIND can be simulated interactively with MAgPIE, a global land-use allocation model that is further connected with the dynamic vegetation model LPJmL, to ensure consistent projections of the price and quantity of bioenergy and GHG emissions. REMIND-MAgPIE has been extensively applied in numerous studies, assessments, and international projects, contributing a marker scenario to the Intergovernmental Panel on Climate Change (IPCC) Assessment Reports and providing the largest number of scenarios in the IPCC Sixth Assessment Report (AR6) Scenario Database among IAMs (Riahi et al. 2022; Peters et al. 2023).

This study utilizes REMIND-MAgPIE (versions 3.2.0-4.8.0), as developed for the Horizon Europe projects RESCUE and OptimESM, to generate emissions scenarios up to 2100 for use in ESM simulations. Within this modeling framework, five different CDR techniques are represented: afforestation, BECCS, DACCS, industry CCS, and OAE. Other CDR techniques, such as enhanced weathering (Goll et al. 2021; Gaucher et al. 2023), are not included in the current version of REMIND-MAgPIE. The baseline scenario assumes a middle-of-the-road pathway (Shared Socioeconomic Pathway 2 (SSP2)) (Dellink et al. 2017).

At the time of writing the manuscript, there are two sets of mitigation scenarios generated from REMIND-MAgPIE in these Horizon Europe projects. First, the Tier-1 scenarios, which are designed for ESM simulations, consist of four scenarios until 2100 to limit the end-of-century warming to 1.5 °C and 2 °C (corresponding to the cumulative carbon budgets of 500 and 1,150 $GtCO_2$, respectively) with high and low temperature overshoot. The high and low overshoot trajectories are generated through the end-of-century budget and peak budget approaches, respectively (Riahi et al. 2021). Second, and more central to this study, are the Tier-1 sensitivity scenarios, that is, ten scenarios generated under a wider range of cumulative carbon budgets (400, 450, 500, 600, 750, 1000, 1300, 1700, 2300, and 3000 $GtCO_2$) using the end-of-century budget approach. All sensitivity scenarios are assumed to follow the 2030 targets outlined in the Nationally Determined Contributions (NDCs). From 2030 onwards, abatement and removal levels are endogenously calculated within the model, following globally uniform carbon price pathways that are calculated exogenously based on the Hotelling rule (Hof et al. 2021) in accordance to respective cumulative carbon budgets. While not directly simulated by ESMs, these sensitivity scenarios serve as inputs for developing and calibrating emIAM, as described in the subsequent section. We selected the end-of-century budget scenarios for emulator development because they are primarily driven by carbon price, unlike the peak budget scenarios, which incorporate additional non-price-driven constraints to force net-zero $CO_2$ emissions.

## 2.2. emIAM

Our approach to emulating IAMs builds upon the framework established by (Xiong et al. 2025) (thereafter, X25), which utilizes MAC curves to represent the economic response of complex IAMs to varying carbon prices. Such use of MAC curves has its roots in earlier studies (Nordhaus 1991; Ellerman and Decaux 1998; O'Neill 2003; van Vuuren et al. 2004; Johansson et al. 2006; Tanaka et al. 2013; Su et al. 2017). We extend this framework by developing an IAM emulator specifically tailored to REMIND-MAgPIE, incorporating model-specific insights gained through collaboration with the model developers, and explicitly accounting for sectoral details of GHG emissions and various CDR approaches. Our primary objective is to apply this emulator to generate long-term global GHG and CDR scenarios extending to 2300 under a range of climate strategies to provide a basis for further explorations by ESMs. While X25 attempted to expand their IAM emulator to regional dimensions, the present study focused on developing a global-level emulator. An overview of the emulation approach is illustrated in Fig. 1.

### *2.2.1. Data used to Extract MAC Curves from REMIND-MAgPIE*

Our IAM emulator was trained using the set of REMIND-MAgPIE sensitivity scenarios as described earlier, specifically those generated prior to the scenario harmonization to adjust with historical emissions (Gidden et al. 2019) (Fig. 2). We utilized the unharmonized scenarios because emIAM aims to capture the fundamental, price-driven behavior of the IAM, isolated from the effects of scenario harmonization. From this set of unharmonized scenarios, we extracted the relationships between the imposed carbon price and the resulting emission abatement (in percentage reduction relative to the baseline for each gas and sector) or the removal level (in $GtCO_2$ for each CDR technique). The baseline scenario used for these calculations accounts for current policies but not NDCs. The price-quantity relationships were individually derived for each of the three major GHGs ($CO_2$, $CH_4$, and $N_2O$) originating from each of the two sectors (energy-related and non-energy-related sectors), as well as for each of the five CDR methods (afforestation, BECCS, DACCS, industrial CCS, and OAE). The energy-related sector includes emissions from "Industrial Sector," "Transportation Sector," "Aircraft," "International Shipping," "Residential Commercial Other," and "Solvents Production and Application" in the original emission database. The non-energy-related sector includes emissions from "Aggregate - Agriculture and LUC" and "Waste." Climate forcers other than the three major GHGs were not incorporated in our IAM emulator because their emissions generally do not respond to carbon price, rather driven primarily by other factors such as clean air policies (Supplementary Fig. 1). A notable exception is ammonia emissions, which respond to carbon price significantly, as they are emitted from agriculture; however, we did not incorporate ammonia in our emulator because it is not an input to the ESMs considered in the Horizon Europe projects.

#### *2.2.2. MAC Curves: Definition and Functional Forms*

The extracted price-quantity relationships are commonly referred to as MAC curves. A MAC curve generally depicts the marginal cost, or willingness to pay, for abating an additional unit of emission relative to a given reference level (Kesicki and Strachan 2011). MAC curves are typically derived from various databases representing a specific point in time (McKinsey & Company 2013; Harmsen et al. 2019; Goldman Sachs 2023). These data-derived MAC curves are thus considered as being contingent on the chosen period. Model-derived MAC curves, such as those developed in this study, can also exhibit temporal variations arising from dynamic processes and constraints within the IAM, such as capital stock turnover, technological learning, and limits on the rate of technology deployment (Krey et al. 2019).

However, our analysis revealed that the price-quantity relationship obtained from REMIND-MAgPIE can be approximated by a single, time-independent MAC curve for most cases, at least during the latter half of the 21st century, with the exception of non-energy-related $CO_2$ emissions (Fig. 3). This finding suggests that the MAC curves derived from REMIND-MAgPIE are *approximately* time-independent within the ranges and scopes noted above. The observed time-independency of the REMIND-MAgPIE MAC curves during the latter half of this century is consistent with the behavior of several, but not all, IAMs, as discussed extensively in X25. While X25 tested time-dependent MAC curves for IAMs that exhibited strong temporal variations in their price-quantity relationships (Section 5 of X25), this study maintains the assumption of time-independency for the MAC curves, given the emulator's main application of extending scenarios beyond the traditional time horizon. That is, for our REMIND-MAgPIE emulator, the observed time-independency between 2050 and 2100 provides a foundation for extending the use of these MAC curves beyond 2100 to develop very long-term scenarios.

To represent the relationships between carbon price and abatement/removal levels, our MAC curves employ one of the following two functional forms:

$$f(x) = a \times x^b + c \times x^d \qquad \text{(used for CDR techniques)} \qquad (1)$$

$$ln\,(f(x) + 1) = a \times x^b + c \times x^d \qquad \text{(used for sectoral GHG emissions)} \qquad (2)$$

where:

- $f(x)$ represents the carbon price (in US$2010 per ton of $CO_2$-equivalent emissions, using the 100-year Global Warming Potential (GWP100) from the IPCC AR5; 28 for $CH_4$ and 265 for $N_2O$).
- $x$ represents either the emission abatement level (in percentage reduction relative to the baseline for each GHG and sector) or the CDR level (in $GtCO_2$ for each CDR).
- $a, b, c,$ and $d$ are fitted parameters, optimized for each GHG, sector, and CDR.

Equation (1) is applied to all five CDR techniques, while equation (2) is applied to $CO_2$, $CH_4$, and $N_2O$ emissions from both the energy-related and non-energy-related sectors. The functional forms are designed to be as simple as possible, yet adequately elaborated to capture the complex relationship between carbon price and abatement/removal levels. The choice between the two functional forms was based on the underlying data distribution and the ability of each form to accurately capture the model-extracted relationship between carbon price and abatement/removal levels. The optimization process involves minimizing the sum of squared errors between the predicted and observed values. The resulting fitted MAC curves are shown in Fig. 3, demonstrating a high goodness of fit, with the coefficients of determination ($r^2$) exceeding 0.9 in most cases. The parameter values of the MAC curves are summarized in Table 1.

### 2.2.3. Dynamic Constraints: Rate and Inertia Constraints

In addition to the MAC curves, emIAM incorporates a few other key elements derived from the REMIND-MAgPIE sensitivity scenarios. These include dynamic constraints that control the extent, rate, and inertia of emissions reductions and CDR deployment. Specifically, we derived "rate constraints," representing the maximum allowable annual change in abatement or removal levels for each GHG, sector, and CDR. These constraints were derived from the 99% range of the first derivative of abatement or removal levels for each GHG, sector, and CDR found in the sensitivity scenarios (Supplementary Fig. 2). Similarly, we derived "inertia constraints," representing the maximum allowable annual change in the *rate* of abatement or removal levels (i.e., the 99% range of the second derivative of the abatement or removal levels) (Supplementary Fig. 3). The use of 99% is arbitrary but works for our purpose, while X25 used the 3-sigma values of data distributions to remove outliers. These rate and inertia constraints capture the limits on how rapidly abatement measures and removal techniques can be deployed and scaled up under factors such as capital stock, infrastructure development, and technological learning.

*2.2.4. Carbon Price Constraints and Treatment of OAE*

We set up an upper bound for the carbon price at 1,600 US$2010 per ton of $CO_2$ emissions, based on the maximum carbon price found in the REMIND-MAgPIE sensitivity scenarios. This upper limit defines the maximum extent of the MAC curves, thereby determining the maximum achievable abatement or removal level for each sectoral GHG emission or CDR technique. Imposing a common upper bound on the carbon price across all sectors and CDRs ensures consistency when the MAC curves are integrated within emIAM and linked through a uniform carbon price (see Section 2.4.3). This approach differs from X25, in which maximum abatement levels were set individually for each gas without a common maximum carbon price. However, the difference in methodology is justified since MAC curves were treated independently in X25.

While emIAM generally treats all abatement and removal options consistently, OAE presents a unique case. Within the ACC2 framework, OAE could be modeled explicitly by simulating the dynamic changes in ocean alkalinity, the associated shifts in the thermodynamic equilibrium of carbonate species, and the resulting changes in pH, thereby influencing the ocean $CO_2$ uptake (Mackenzie and Lerman 2006; Tanaka et al. 2007; Kvale et al. 2012).

However, given our overarching goal of emulating REMIND-MAgPIE, emIAM adopts a similar approach for OAE as for other CDR techniques. This ensures uniformity in the emulator's structure and simplifies its application, while acknowledging the potential for more detailed process-based modeling of OAE within ACC2 in future work.

## 2.3. ACC2

To assess the climate implications of long-term emissions scenarios, we coupled emIAM with the reduced-complexity climate model ACC2 (Tanaka et al. 2007, 2021; Tanaka and O'Neill 2018). ACC2 builds upon earlier generations of reduced-complexity climate models (Hasselmann et al. 1997; Hooss et al. 2001; Bruckner et al. 2003) and consists of four interconnected modules: carbon cycle, atmospheric chemistry, physical climate, and economy. For this study, the original economy module of ACC2 was replaced with the newly developed IAM emulator, emIAM, described in the preceding sections. This update allows for a dynamic interaction between the emission reductions and removals calculated by emIAM and the resulting climate outcomes simulated by ACC2. While the original economy module, with three aggregated MAC curves for total $CO_2$, $CH_4$, and $N_2O$ emissions (Tanaka and O'Neill 2018), allows ACC2 to calculate least-cost pathways for a given temperature target, emIAM, with eleven MAC curves for sectoral GHG emissions and CDR techniques as exhibited from the latest REMIND-MAgPIE, provides a more updated and detailed representation of climate mitigation involving the energy-land-economy system.

The ACC2 model individually calculates the temperature contributions of various gases and aerosols, including $CO_2$, $CH_4$, $N_2O$, 29 species of halocarbons, $SF_6$, tropospheric and stratospheric $O_3$, stratospheric water vapor, and aerosols (the direct effect of sulfate aerosols, the direct effect of black carbon and organic aerosols, and the indirect effects of all aerosols). ACC2 individually evaluates the radiative forcing of each climate forcer, without aggregating using emission metrics such as GWP100, the use of which does not ensure temperature equivalence upon emission aggregation (Allen et al. 2022; Mastropierro et al. 2023). The physical climate module is based on an energy balance and heat diffusion model DOECLIM (Kriegler 2005). The atmospheric chemistry module accounts for OH chemistry to compute $CH_4$ and tropospheric $O_3$ concentrations, taking into account emissions of $NO_x$, CO, and VOC. The carbon cycle module is a box model comprising three ocean boxes, a coupled atmosphere-mixed layer box, and four land boxes. It incorporates key processes such as the saturation of ocean $CO_2$ uptake due to changes in carbonate chemistry and the enhancement of land $CO_2$ uptake through $CO_2$ fertilization as parameterized by the beta factor. Climate-carbon cycle feedbacks are also included, parameterized using a Q10 formulation. Following the best estimate from the IPCC AR6, we adopted an equilibrium climate sensitivity of 3°C for a doubling of atmospheric $CO_2$ concentration, within the assessed range of 2.5 to 4.0°C (IPCC 2023). Other uncertain model parameters were calibrated using an inverse modeling approach (Tarantola 2005), fitting the model to historical data considering model priors since 1750 (Tanaka et al. 2009). The model is written in GAMS and numerically solved using CONOPT3 and CONOPT4, nonlinear optimization solvers included in the GAMS software package.

ACC2 has a well-established history, having participated in several model intercomparison projects (Joos et al. 2013; Nicholls et al. 2020). Its model responses have been evaluated and compared with those of other models (Xiong et al. 2022; Mastropierro et al. 2023; Melnikova et al. 2023a). Furthermore, ACC2 has been applied in many studies analyzing pathways to achieve the Paris Agreement targets (Tanaka and O'Neill 2018; Tanaka et al. 2021), including assessments of GHG removal options (Gaucher et al. 2023, 2025). For this study, we benchmarked the performance of ACC2 (without the economic module) against other reduced-complexity climate models including MAGICC and FaIR, and OSCAR using commonly used scenarios RCP2.6 and RCP4.5 in emission-driven mode (Nicholls et al. 2020). As shown in Fig. 4 and Supplementary Fig. 4 (for RCP2.6 and RCP4.5, respectively), the model response of ACC2 is generally consistent with those of other reduced-complexity climate models. Notably, the long-term $CO_2$ concentration projections of ACC2 are comparable to those of OSCAR. This similarity in long-term $CO_2$ concentration projections is particularly important for the development of our extended scenarios, as concentration data generated from OSCAR using emission data from emIAM-ACC2 will be provided to ESMs for concentration-driven simulations in the RESCUE project.

## 2.4. Evaluation of emIAM – approach and key considerations

Before applying emIAM to generate extended scenarios to 2300, we evaluated its ability to reproduce the original REMIND-MAgPIE scenarios under comparable conditions. This evaluation process (summarized in Table 2) involved a series of experiments designed to incrementally assess the emulator performance under increasingly complex constraints across multiple dimensions. These dimensions are detailed below in three parts: i) target types, ii) scope of emission evaluations, and iii) uniform carbon price and GWP100. Due to the limited number of available REMIND-MAgPIE scenarios, the same set of sensitivity scenarios was used for both emulator development and evaluation. However, to further probe the emulator performance, we conducted an additional evaluation using the Tier-1 scenarios with high temperature overshoot. The experiments conducted in this study build on those developed in X25, with further elaborations to investigate the assumptions associated with uniform carbon price and GWP100 and to delve more deeply into the role of the carbon price pathway as a key determinant of emulator performance. The results of the series of tests are presented in Figs. 5 to 7.

### *2.4.1. Target Types for Emulator Evaluation*

Our experiments used three primary types of targets: i) carbon price pathway, ii) cumulative carbon budget, and iii) temperature.

- *Carbon Price Pathway*: When a specific carbon price pathway is prescribed, the corresponding emission pathway is uniquely determined within the emulator without any optimization. This experiment isolates the impact of any deviations between the fitted MAC curve and the original price-quantity data from REMIND-MAgPIE.
- *Cumulative Carbon Budget*: When a cumulative carbon budget is imposed, the emulator dynamically allocates emissions over time by minimizing the total discounted abatement costs (assuming a 5% discount rate). This allocation is primarily driven by the exponentially rising carbon price pathway (consistent with the Hotelling rule), modulated by the MAC curves and the dynamic constraints (i.e., rate and inertia constraints, as well as upper limits on abatement and removal). Importantly, experiments with carbon price pathways and cumulative carbon budgets involve only emIAM, without utilizing the carbon cycle, atmospheric chemistry, and physical climate modules of ACC2. Therefore, these experiments serve as a direct evaluation of the fidelity of emIAM.
- *Temperature*: When a temperature target is imposed, the emulator accounts for the different physical and biogeochemical characteristics of GHGs within the Earth system as represented in ACC2 in allocating emissions over time. This introduces complexities not present in REMIND-MAgPIE, which relies on the concept of cumulative carbon budget. The distinct and complex responses of $CO_2$, $CH_4$, and $N_2O$ to temperature targets cannot be adequately captured by the static GWP100 typically used to convert emissions within IAMs. Therefore, experiments using temperature targets represent an application of the coupled emIAM-ACC2, while also providing an additional, albeit indirect, evaluation of the fidelity of emIAM. Note that the temperature target is based on the temperature data calculated with ACC2, not those with MAGICC (available in the original REMIND-MAgPIE scenario database), to avoid introducing model differences.

*2.4.2. Scope of Emissions for Emulator Evaluation*

The emulator was initially tested for its ability to reproduce individual sectoral $CO_2$ emission and CDR pathways. Subsequently, it was tested for reproducing the total anthropogenic emissions of each GHG ($CO_2$, $CH_4$, and $N_2O$) individually, thus reflecting trade-offs among different sectors and CDRs within a single gas. Finally, it was tested for reproducing the total emissions of all three GHGs simultaneously, thereby addressing trade-offs among different gases. When calculating least-cost emission pathways for only one GHG at a time (e.g., $CO_2$), we constrained the emissions of the other GHGs ($CH_4$ and $N_2O$) to follow the respective pathways from the corresponding REMIND-MAgPIE sensitivity scenarios. A simultaneous calculation of all three GHGs requires a $CO_2$-equivalent budget target using GWP100 for emission aggregation. For aggregating GHG emissions, we adopted GWP100 as the default metric without delving into metric issues (Reisinger et al. 2022), while noting that the cumulative $CH_4$ budget based on GWP100 does not offer any meaningful physical interpretation (Mastropierro et al. 2023), whereas the cumulative $CO_2$ budget is an indicator of the global-mean temperature change (Matthews et al. 2009).

*2.4.3. Uniform Carbon Price and GWP100*

We investigated the impact of the following two key assumptions on emulator performance: (1) the use of a uniform carbon price across sectors and CDRs, and (2) the use of GWP100 to fix the price ratios between different GHGs at their respective metric values (i.e., 28 for $CH_4$ to $CO_2$ and 265 for $N_2O$ to $CO_2$, as in the IPCC AR5). The first assumption constrains the relationships among the abatement and removal levels of different sectors and CDRs, while the second assumption constrains those of different gases (Tanaka et al. 2021). Both of these constraints are used in REMIND-MAgPIE; however, they are not necessarily always binding for all abatement and removal measures, depending on the states of other constraints and the way individual abatement and removal measures are represented (Supplementary Fig. 5). Examples in REMIND-MAgPIE include:

- In high emissions scenarios, $CO_2$ emissions from fossil fuel combustion can decrease later in this century due to rising costs of extracting fossil fuels (Mohr et al. 2015), which is not directly driven by the carbon price.
- $CH_4$ emission reductions associated with clean air policies that are assumed to be progressively implemented in a differentiated manner around the world reflecting the state of development (Amann et al. 2020) are not dynamically linked to the carbon price, even though they are introduced in accordance with cumulative carbon budgets in REMIND-MAgPIE.
- Some $CH_4$ emission reduction measures, such as those reducing $CH_4$ leakage from gas-fired power plants during venting (Lauvaux et al. 2022), are driven by the carbon price (or by the $CH_4$ price, which is linked to the carbon price multiplied via GWP100). On the other hand, other $CH_4$ emission reduction measures, such as those incentivising an early retirement of coal-fired power plants (Tanaka et al. 2019), are only indirectly driven by the carbon price, as they occur as a co-reduction of $CO_2$ mitigation measures.

By comparing the emulator output with and without these assumptions, we assess their influence on the reproduced sector-, CDR-, and gas-specific emission pathways to understand which set of assumptions best reproduces the original REMIND-MAgPIE scenarios.

## 3. Results

### 3.1. Overview of the scenarios

We generated a suite of extended scenarios to 2300 using the coupled emIAM-ACC2 modeling framework described in the preceding section. The methods to derive the extended baseline and mitigation scenarios are summarized in Table 3. An overview of these extended scenarios is presented in Fig. 8, with further details of selected emission pathways by gas, sector, and CDR technique illustrated in Figs. 9 to 12 and those of all others in Supplementary Figs. 6 to 11.

#### *3.1.1. Prescribed Pre-2100 Scenarios*

To maintain consistency with the scenarios used for ESM simulations up to 2100, we prescribed a set of harmonized scenarios to the emulator for the period up to 2100. This set of scenarios include the four Tier-1 scenarios, characterized by cumulative carbon budgets of 500 and 1,150 $GtCO_2$ and based on both end-of-century and peak budget approaches. It also includes a subset of sensitivity scenarios with cumulative carbon budgets of 2,300 and 3,000 $GtCO_2$ (using the end-of-century budget approach) and the baseline scenario assuming only current policies (Fig. 8). This diverse range of pre-2100 scenarios allows for the extension of scenarios exhibiting various overshoot profiles. While using harmonized scenarios for the pre-2100 period might appear inconsistent with the use of unharmonized scenarios for emulator development, this discrepancy is negligible because harmonization primarily affects the scenarios in the first half of the 21st century.

#### *3.1.2. Scenario Extension to 2300: Temperature Targets and Overshoot Profiles*

With these harmonized scenarios prescribed up to 2100, the coupled emIAM-ACC2 was used to extend the scenarios to 2300 by calculating least-cost pathways for specified temperature targets (2, 1.5, and 1 °C) applied from respective target years. In this approach, the temperature is allowed to exceed the target level before the target year is reached, since it is less costly to invest in mitigation later in the cost-effective optimization framework (Emmerling et al. 2019). Consequently, the length and the magnitude of overshoot are determined by the cost minimization performed in emIAM-ACC2, reflecting the abatement and removal constraints emulated from REMIND-MAgPIE, the inertia of Earth system processes represented in ACC2, and the assumed discount rate (5%).

#### *3.1.3. Net-zero $CO_2$ Emissions and other Emissions Constraints*

To ensure that the extended scenarios are useful for ESM simulations, we imposed a constraint of net-zero $CO_2$ emissions from the year that the specified temperature target is applied. Similarly to the Zero Emissions Commitment Model Intercomparison Project (ZECMIP) (Jones et al. 2019), this scenario design allows ESMs to investigate the response to zero $CO_2$ emissions following historical and future pathways. After net-zero $CO_2$ emissions are imposed (and the temperature target is reached), the temperature slightly declines over time. This reflects the model dynamic of ACC2, which is characterized by a negative ZEC (Mastropierro et al. 2023). The ZEC parameters from multiple models generally show significant variations with slightly negative means (MacDougall et al. 2020). This suggests that the temperature response of EMSs to zero $CO_2$ emissions combined with positive non-$CO_2$ emissions in our extended scenarios can be highly model-dependent. Another reason why the temperature decreases without non-$CO_2$ emissions increase is the assumption on emissions discussed below.

Furthermore, we assumed that energy-related and non-energy-related emissions would not increase after 2100. This assumption facilitates the development of long-term pathways that converge towards the Paris Agreement temperature targets without introducing unnecessary complexity in the post-2100 pathways (i.e., avoiding emission recoveries). Importantly, this non-increasing emission assumption is not applied for CDRs, which means that the deployment of CDRs can be not only scaled up but also scaled down in the extended scenarios. The CDR deployment, including the potential for phase-out, serves as a key driver in shaping the temperature overshoot profiles.

A limitation of this approach is the lack of data to constrain the rate and inertia parameters for scaling down CDR deployment, as IAMs typically do not simulate decreasing CDR deployment. Therefore, we made a simplifying assumption that the rate and inertia constraints for scaling down CDRs are of the same magnitudes but opposite in sign to those derived from REMIND-MAgPIE for scaling up. From a scenario development perspective, this assumption can be varied to explore the implications of different CDR phase-out strategies.

#### *3.1.4. Development of the Extended Baseline Scenario*

Extending scenarios beyond 2100 necessitates an extended baseline scenario. Due to the scarcity of literature on multi-century baseline projections and the inherent difficulty in projecting socio-economic factors over such long time scales, we developed our extended baseline scenario based on a set of simplifying assumptions, as detailed below and illustrated in Supplementary Fig. 6. The design of the extended baseline scenario also informs the range of applicability for the emulator, as emIAM was constructed based on a relatively high baseline scenario simulated by REMIND-MAgPIE until 2100.

#### *3.1.5. Details of the Extended Baseline and Mitigation Scenarios*

- *Energy-related $CO_2$ Emissions*: The extended baseline scenario for energy-related $CO_2$ emissions was developed based on the economic extractability of available fossil fuel reserves, using output from REMIND-MAgPIE. This approach indicates that oil and coal will become economically unextractable by approximately 2200, while economically accessible natural gas reserves may persist longer despite a small amount (Supplementary Fig. 6). This led us to assume that by 2200, economically recoverable fossil fuel reserves are practically depleted, suggesting zero energy-related $CO_2$ emissions from this point onward. By 2150, the extended baseline scenario for energy-related $CO_2$ emissions falls to approximately 75% of the 2020 baseline level. Because our MAC curves, which rely on percentage reductions from the baseline, become less reliable as the baseline deviates from the calibrated range and eventually approaches zero, we employ the emulator only until 2150 for energy-related $CO_2$ emissions. Beyond 2150, we apply a linear assumption to approximate the declining baseline projection using resource assumptions in REMIND-MAgPIE toward zero emissions in 2200 and scale the extended mitigation scenarios for energy-related $CO_2$ emissions proportionally with the extended baseline scenario. This implies that all energy-related $CO_2$ emissions reach zero by 2200.

- *Non-energy-related $CO_2$ Emissions*: The extended baseline scenario for non-energy-related $CO_2$ emissions was assumed to decline linearly to zero by 2150 and remain at zero thereafter, reflecting limits in land availability for further emissions. To be consistent with this assumption, the emulator approach driven by carbon price was not applied to non-energy-related $CO_2$ emissions. The extended mitigation scenarios for non-energy-related $CO_2$ emissions are assumed to decline linearly to zero by 2150. This approach aligns with the methodology of (Meinshausen et al. 2020) and is also consistent with our spatial extension approach, which assumes that spatial emission and forcing patterns are held unchanged after 2100.
- *CDRs*: The extended baseline scenario assumes zero deployment of all CDR techniques, except afforestation. Afforestation is deployed at a significant scale in the pre-2100 baseline scenario (approximately 1.2 GtC/year in 2100, Supplementary Fig. 5). To maintain consistency with the assumption for non-energy-related $CO_2$ emissions, both the extended baseline and mitigation scenarios for afforestation are assumed to decline linearly to zero by 2150. For the rest of CDRs, the emulator approach was applied over the entire time horizon.

- *Energy-related Non-$CO_2$ GHGs*: The extended baseline scenarios for energy-related $CH_4$ and $N_2O$ emissions were scaled proportionally with the extended baseline for energy-related $CO_2$ emissions after 2150. Similarly, the extended mitigation scenarios for energy-related $CH_4$ and $N_2O$ emissions were scaled with their respective extended baseline scenarios. Consequently, both baseline and mitigation scenarios for energy-related $CH_4$ and $N_2O$ emissions are projected to approach zero by 2200.
- *Non-energy-related Non-$CO_2$ GHGs*: The extended baseline scenarios for non-energy-related $CH_4$ and $N_2O$ emissions were assumed constant after 2100, reflecting the difficulty in abating emissions related to food production. To derive extended baseline scenarios for these emissions, the MAC curve approach was applied over the entire time horizon.

### *3.1.6. Treatment of Uniform Carbon Price and GWP100 in the Extended Scenarios*

As previously discussed, the REMIND-MAgPIE model, which forms the basis for our emulator, employs a globally uniform carbon price across sectors and regions in the sensitivity scenarios (using the end-of-century budget approach) and utilizes GWP100 to determine the relative prices of $CH_4$ to $CO_2$ and of $N_2O$ to $CO_2$. Fully replicating these assumptions within the emulator during the scenario extension phase (beyond 2100) presents challenges. Therefore, we adopted a modified approach, as detailed below.

*Carbon Price Convergence*: The carbon price pathways inversely calculated from the sectoral emissions and CDR levels using respective MAC curves are approximately consistent across sectors and CDRs for the pre-2100 unharmonized scenarios. However, the consistency is lower in the pre-2100 harmonized pathways due to the effect of harmonization. To ensure consistency across sectors and CDRs in the extended scenarios, we implemented a convergence period. Specifically, we assumed an arbitrarily determined 10-year period beginning in 2100 during which carbon prices are allowed to converge within the cost minimization calculations in em-IAM-ACC2. This transition period provides sufficient time for different sectors and CDRs to adjust their respective carbon prices before a uniform carbon price is enforced from 2110 onwards. While the carbon prices are consistent from 2110 onwards in principle, however, there are two notable exceptions, as outlined below.

- Non-energy-related $CO_2$ emissions and afforestation are excluded from the uniform carbon price regime from 2100 onwards. This is because their pathways are predetermined to decline to zero by 2150 due to factors unrelated to carbon price, such as land availability constraints.
- Energy-related $CO_2$ emissions are excluded from the uniform carbon price regime from 2150 onwards. This reflects the projected depletion of economically recoverable fossil fuel reserves by 2200, rendering further emission reductions independent of carbon price.

Consequently, after 2150, only individual CDRs (excluding afforestation) are subject to a uniform carbon price within the emulator.

*Treatment of GWP100*: GWP100 was used to determine the prices of energy-related $CH_4$ and $N_2O$ emissions relative to energy-related $CO_2$ emissions until 2150, consistent with the period when the emulator approach was applied to these emissions. Beyond 2150, however, we discontinued the use of GWP100 for these emissions because these emissions were assumed to scale proportionally with energy-related $CO_2$ emissions, not directly linked to carbon price. For non-energy-related $CH_4$ and $N_2O$ emissions, GWP100 was not applied, primarily due to the assumption of non-increasing emissions. Applying GWP100-constrained price ratios to non-energy-related $CH_4$ and $N_2O$ emissions would create unnecessary constraints on CDRs (excluding afforestation), which are allowed to scale both up and down in the extended scenarios.

## 4. Discussions

This study introduced a novel emulator-based approach for generating internally consistent scenarios extending to 2300, providing a new basis for exploring long-term climate pathways using ESMs. Our approach leverages the detailed, process-based insights embedded within the pre-2100 scenarios generated by REMIND-MAgPIE and extends them into the period up to 2300. This contrasts with previous approaches to scenario extension, as well as those planned for CMIP7, which primarily rely on linearly declining emissions with limited connection to the underlying dynamics of the pre-2100 pathways.

Developing the extended scenarios presented several challenges. Notably, the extended baseline scenario, developed using a set of simplifying assumptions due to the scarcity of multi-century socioeconomic projections, showed significantly lower emissions than the pre-2100 baseline scenario from REMIND-MAgPIE. This discrepancy led to certain limitations to the range of emulator applications and necessitated several adjustments to our methodology to develop extended scenarios, as detailed in previous sections. It also highlights the inherent uncertainties associated with projecting socio-economic development over such long time horizons. Despite these challenges, we strived to minimize the number of additional assumptions and maintain internal consistency within the emulator, as well as coherency with REMIND-MAgPIE, in the treatment of different GHGs, sectors, and CDRs over the entire time horizon.

The IAM emulator, emIAM, developed in this study, is specifically tailored to REMIND-MAgPIE through close collaboration with the model developers. The extensive testing and validation process provides a clear understanding of its applicability and potential biases. emIAM is far lighter than REMIND-MAgPIE in terms of computational burden. emIAM could be used independently to produce pathways until 2100, with only a notable caveat on non-energy-related $CO_2$ emissions in our current approach. emIAM is also versatile, as it can be coupled to other reduced-complexity climate models. Although this emulator approach may be adaptable to other IAMs, as suggested by X25, careful consideration of model-specific characteristics and limitations is crucial. One specific limitation of our current emulator is the difficulty in accurately reproducing carbon price pathways when the emulator is used for a carbon budget or temperature target. This could potentially be improved by incorporating additional constraints or parameters that more closely reflect the internal dynamics of REMIND-MAgPIE.

The extended scenarios generated in this study are primarily intended to serve as inputs for ESM simulations and subsequent climate change impact assessments. They may also provide a valuable scenario framework for investigating tipping elements. While many of the scenarios refer to temperature targets aligned with the Paris Agreement, it is crucial to emphasize that these scenarios are not intended to provide direct policy prescriptions. Furthermore, the large temperature overshoots exhibited in some scenarios, which rely on massive deployment of CDRs, should be interpreted cautiously. This study did not explicitly assess the potential side effects of large-scale CDR deployment on other sustainability goals, such as food security and land use, over the extended time horizon.

In conclusion, this work represents a significant step towards developing more robust and consistent long-term climate scenarios. The emulator-based approach offers a valuable tool for bridging the gap between IAMs and ESMs and for exploring the complex interactions between human activities and the Earth system processes over centennial and multi-centennial timescales. Further development and application of this approach, including refinement of the emulator, exploration of alternative baseline scenarios, and expansion of the scenario set, will be useful for guiding future research on long-term climate change mitigation.

**Competing interests** The authors declare that they have no competing interests.

**Supplementary Information** Supplementary information for this article is available.

| Gas | *a* | *b* | *c* | *d* | MaxABL | Max1st | Max2nd |
|---|---|---|---|---|---|---|---|
| CO2 | 5.12E-03 | 1.41E+01 | 5.27E+00 | 4.54E-01 | 146.4 | 5.5 | 2.0 |
| CH4 | 7.70E+00 | 5.47E-01 | 1.75E+01 | 4.61E+00 | 59.4 | 2.9 | 0.9 |
| N2O | 1.16E+01 | 7.83E-01 | 0.00E+00 | 0.00E+00 | 56.2 | 2.3 | 1.0 |
| CO2ENG | 5.43E+00 | 4.06E-01 | 4.10E+00 | 5.03E+01 | 98.6 | 5.7 | 1.1 |
| CH4ENG | 5.27E+00 | 4.73E-01 | 2.23E+00 | 4.46E+01 | 99.9 | 6.1 | 1.5 |
| N2OENG | 1.58E+06 | 1.00E+02 | 5.57E+00 | 2.50E-01 | 87.3 | 4.9 | 1.7 |
| CO2NEG | 9.54E+00 | 1.19E+00 | 1.25E+20 | 1.26E+02 | 69.3 | 4.5 | 1.1 |
| CH4NEG | 7.26E+00 | 4.72E+00 | 1.04E+01 | 6.12E-01 | 52.5 | 2.3 | 0.9 |
| N2ONEG | 1.02E+01 | 4.97E-01 | 1.03E+01 | 4.38E+00 | 46.9 | 2.3 | 0.5 |
| CDR Total | 2.25E-05 | 1.00E+01 | 1.76E+02 | 3.64E-01 | 21.840 | 0.728 | 0.143 |
| CDR Afforestation | 6.94E+03 | 3.41E+00 | 2.49E+02 | 4.21E-01 | 2.290 | 0.041 | 0.010 |
| CDR BECCS | 1.83E-07 | 2.54E+01 | 2.37E+02 | 5.28E-01 | 8.924 | 0.294 | 0.083 |
| CDR DACCS | 5.18E-02 | 1.27E+01 | 5.29E+02 | 3.49E-01 | 7.907 | 0.440 | 0.128 |
| CDR Industry | 8.00E+05 | 3.99E+00 | 3.63E+02 | 2.84E-01 | 0.744 | 0.046 | 0.011 |
| CDR OAE | 1.57E-08 | 8.03E+01 | 4.80E+02 | 5.15E-01 | 5.000 | 0.297 | 0.089 |

**Table 1 | Parameter values of the MAC curves for sectoral $CO_2$, $CH_4$, and $N_2O$ emissions and CDR techniques and their upper limits on abatement or removal levels, as well as their rate and inertia constraints, as derived from the REMIND-MAgPIE sensitivity scenarios.** See equations (1) and (2) for parameters $a, b, c,$ and $d$. MaxABL denotes the maximum abatement or removal level (% and $GtCO_2$, respectively). The units for $a$ and $c$ are US\$/$tCO_2$. Max1st and Max2nd represent the maximum first (%/year and $GtCO_2$/year) and second derivatives (%/(year)$^2$ and $GtCO_2$/(year)$^2$), respectively, of abatement or removal changes.

| Test name | Target | Emissions to evaluate | Coverage of uniform price* |
|---|---|---|---|
| 1A | Carbon price pathway | Each sector and CDR separately | - |
| 2A | Cumulative $CO_2$ budget from 2020 to 2100 | Each sector and CDR separately | - |
| 2B1 | Cumulative $CO_2$ budget from 2020 to 2100 | Each gas separately ($CO_2$ only) | Energy, non-energy |
| 2B2 | Cumulative $CO_2$ budget from 2020 to 2100 | Each gas separately ($CO_2$ only) | Energy, non-energy, CDR |
| 3C1 | Cumulative GHG budget from 2020 to 2100 | Three gases altogether | Energy, non-energy |
| 3C3 | Cumulative GHG budget from 2020 to 2100 | Three gases altogether | Energy, non-energy, CDR, non-$CO_2$ |
| 4B1 | 2100 Temperature | Each gas separately | Energy, non-energy |
| 4B2 | 2100 Temperature | Each gas separately | Energy, non-energy, CDR |
| **4C1**** | **2100 Temperature** | **Three gases altogether** | **Energy, non-energy** |
| 4C3 | 2100 Temperature | Three gases altogether | Energy, non-energy, CDR, non-$CO_2$ |
| 5C1 | 2100 Temperature and peak temperature | Three gases altogether | Energy, non-energy |
| 5C3 | 2100 Temperature and peak temperature | Three gases altogether | Energy, non-energy, CDR, non-$CO_2$ |

**Table 2 | Experimental design of the tests performed for evaluating the REMIND-MAgPIE emulator.** Test names are coded with three digits. The first digit indicates the simulation target, the second digit the scope of emissions that are evaluated, and the third digit the range of emissions covered by uniform carbon price. The use of GWP100 can incorporate non-$CO_2$ emissions in the regime of uniform carbon price. *Carbon prices applied to individual CDR techniques are assumed to be uniform within CDRs when multiple CDR pathways are simultaneously calculated. **The settings used in Test 4C1 were most akin to those used to derive extended scenarios. See text for details.

| Scenario type | Gas/sector/CDR | Approach |
|---|---|---|
| Extended baseline scenario | Energy-related $CO_2$ | Economic extractability of available fossil fuel reserves based on assumptions used in REMIND-MAgPIE, indicating its peak around 2080, followed by decreasing emissions to zero by 2200 (approximate linear decrease from 2150 to 2200) |
| | Non-energy-related $CO_2$ | Linearly decreasing emissions to zero by 2150 |
| | Energy-related $CH_4$ | Scaled proportionally with energy-related $CO_2$ |
| | Non-energy-related $CH_4$ | Constant at 2100 level |
| | Energy-related $N_2O$ | Scaled proportionally with energy-related $CO_2$ |
| | Non-energy-related $N_2O$ | Constant at 2100 level |
| | CDR Afforestation | Linearly decreasing removals to zero by 2150 |
| | CDR BECCS | Zero |
| | CDR DACCS | Zero |
| | CDR Industry | Zero |
| | CDR OAE | Zero |
| Extended mitigation scenario | Energy-related $CO_2$ | Calculated by the emulator until 2150<br>Linearly decreasing emissions to zero from 2150 to 2200 |
| | Non-energy-related $CO_2$ | Linearly decreasing emissions to zero by 2150 |
| | Energy-related $CH_4$ | Calculated by the emulator until 2150<br>Linearly decreasing emissions to zero from 2150 to 2200 |
| | Non-energy-related $CH_4$ | Calculated by the emulator from 2100 onwards |
| | Energy-related $N_2O$ | Calculated by the emulator until 2150<br>Linearly decreasing emissions to zero from 2150 to 2200 |
| | Non-energy-related $N_2O$ | Calculated by the emulator from 2100 onwards |
| | CDR Afforestation | Linearly decreasing removals to zero by 2150 |
| | CDR BECCS | Calculated by the emulator from 2100 onwards |
| | CDR DACCS | Calculated by the emulator from 2100 onwards |
| | CDR Industry | Calculated by the emulator from 2100 onwards |
| | CDR OAE | Calculated by the emulator from 2100 onwards |

**Table 3 | Methods to derive the extended baseline and mitigation scenarios for each gas, sector, and CDR.** See text for details.

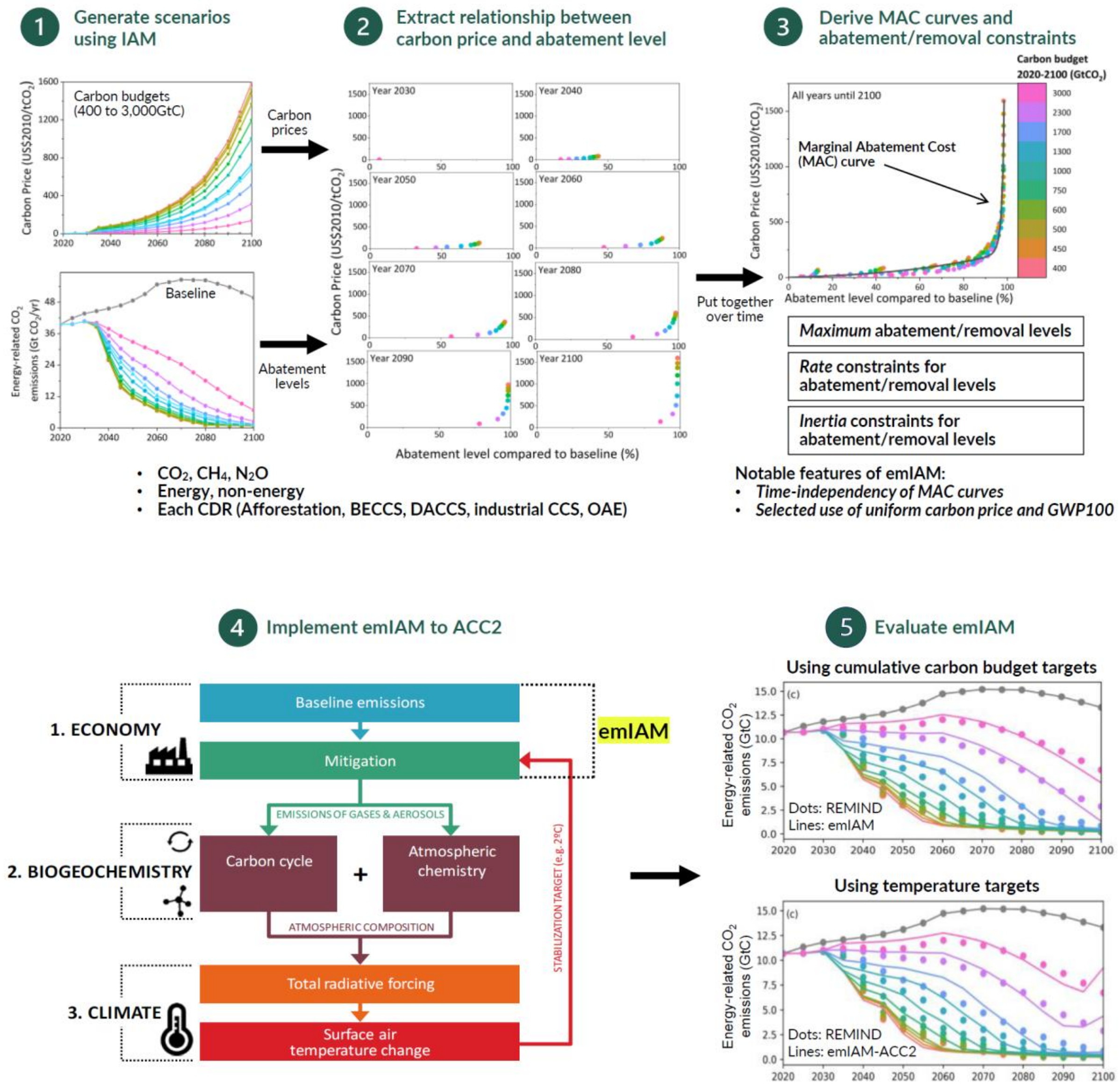

**Fig. 1 | Steps to develop and evaluate the emulator of REMIND-MAgPIE.** The figure above uses the data for global total energy-related $CO_2$ emissions for illustration. Data were obtained from the sensitivity scenarios simulated by REMIND-MAgPIE (versions 3.2.0-4.8.0). The chromatic colors indicate the respective carbon budgets for the period 2019 – 2100 in $GtCO_2$. The gray color indicates the baseline scenario.

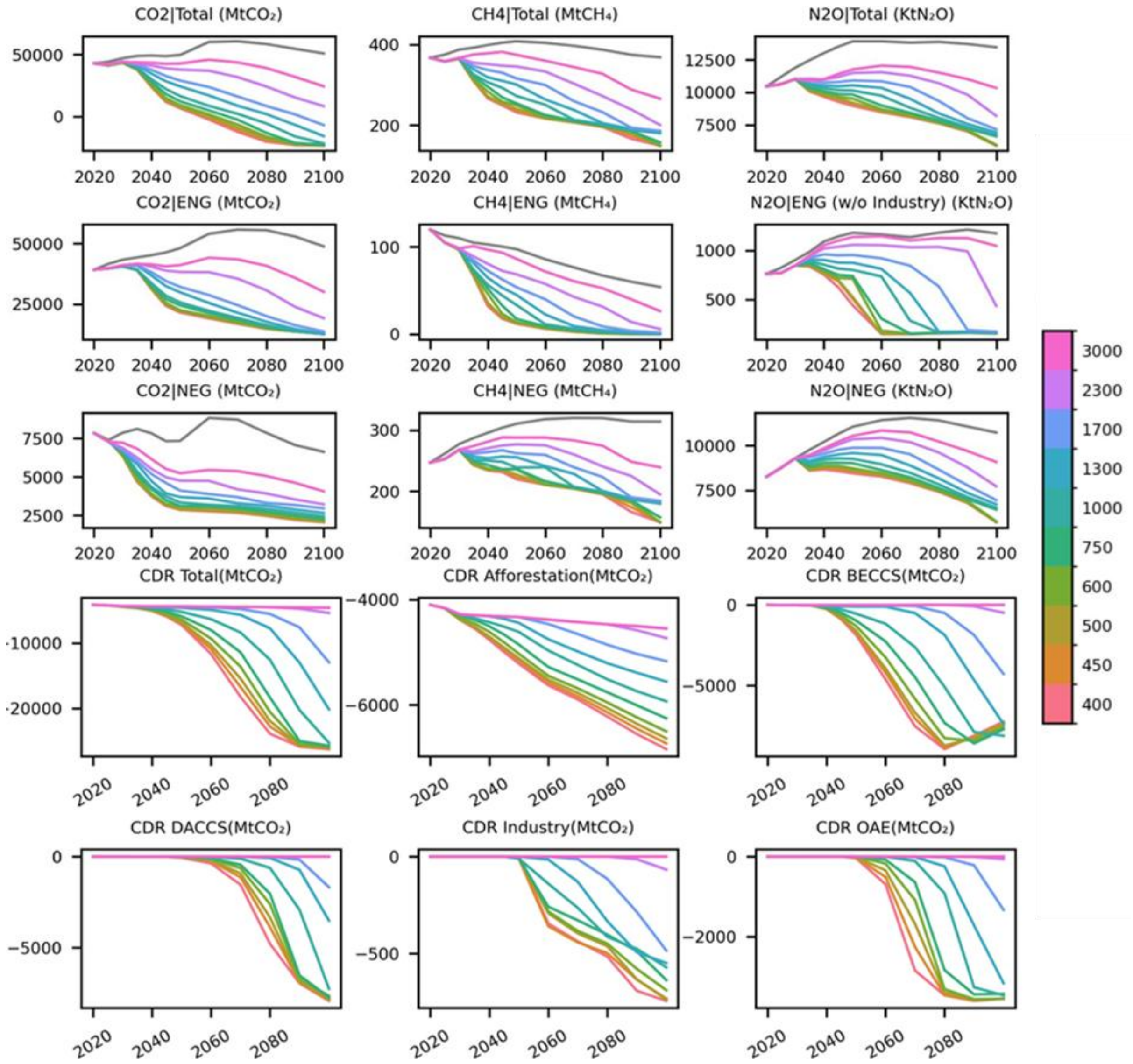

**Fig. 2 | Sensitivity scenarios of three major GHGs used for developing the REMIND-MAgPIE emulator.** The sensitivity scenarios of other climate forcers are shown in Supplementary Fig. 1.

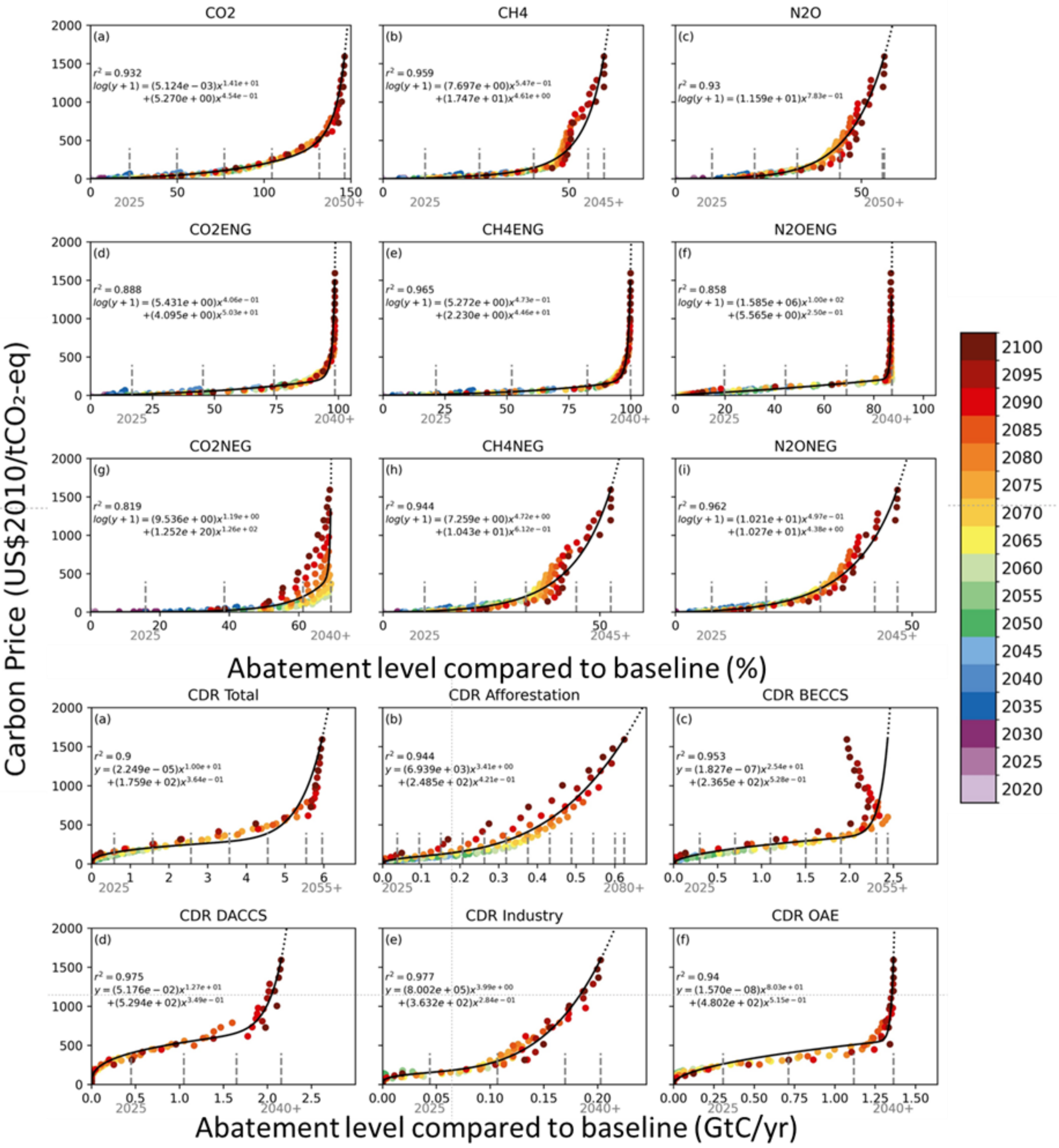

**Fig. 3 | MAC curves by gas, sector, and CDR as extracted from the sensitivity scenarios simulated by REMIND-MAgPIE.**

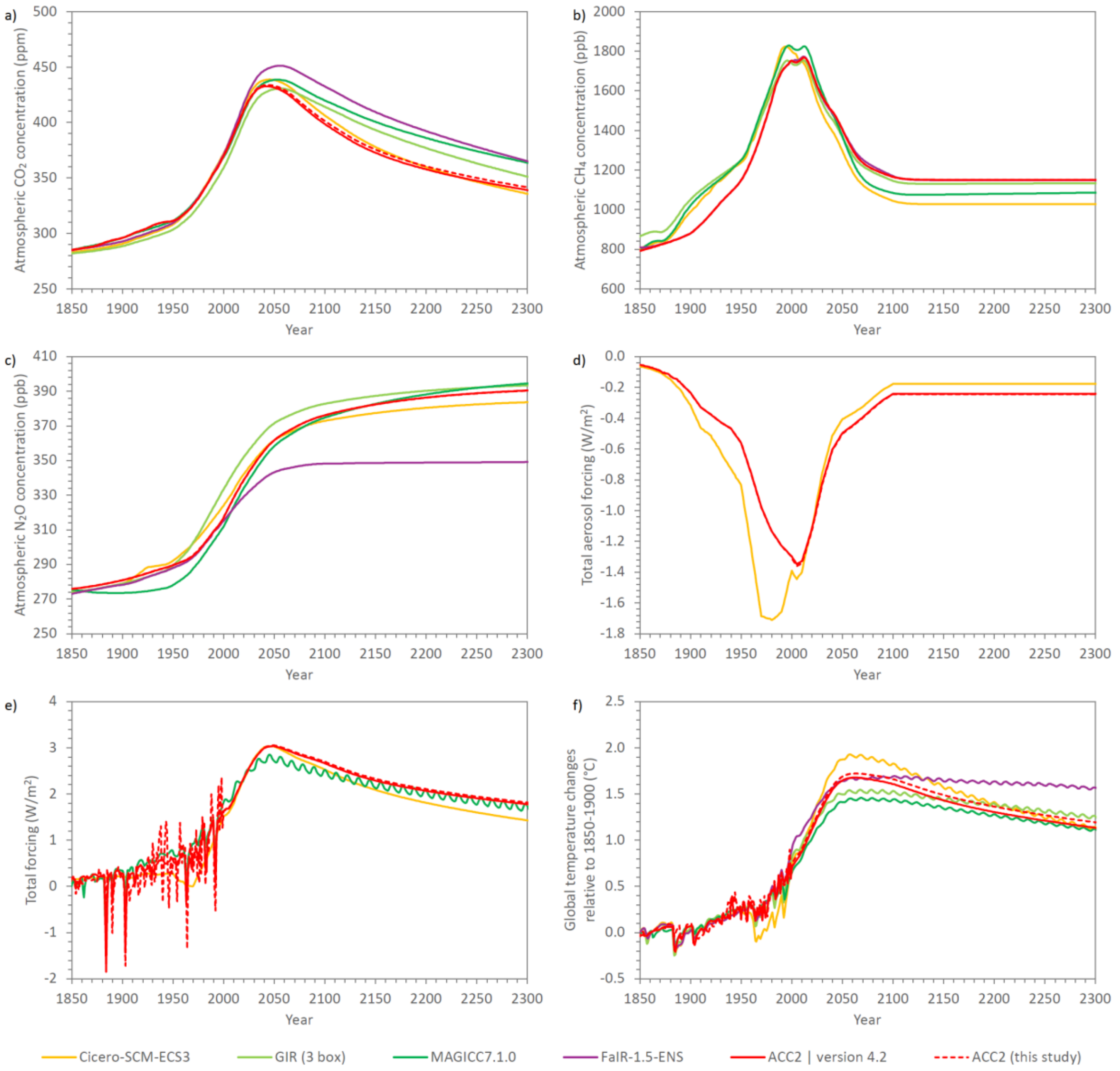

**Fig. 4 | ACC2 and other reduced-complexity climate models simulating the RCP2.6 scenario in the emission-driven mode.** Data are from Nicholls et al. (2020), except for those from the version of ACC2 configured for this study. Results from GHG emission-driven runs are available from the following models as presented: CICERO-SCM-ECS3, GIR (3 box), MAGICC7.1.0, FaIR-1.5-ENS, and ACC2 v4.2. Only available models are shown for each of the following variables: atmospheric $CO_2$, $CH_4$, and $N_2O$ concentration (ppm, ppb, ppb, respectively), total aerosol forcing ($W/m^2$), total forcing ($W/m^2$), and global temperature changes (°C) (panels a to f, respectively). The number of models shown for the radiative forcing terms (panels d and e) is limited because a subset of models report only effective radiative forcing. The temperature data were adjusted for the 1850-1900 reference.

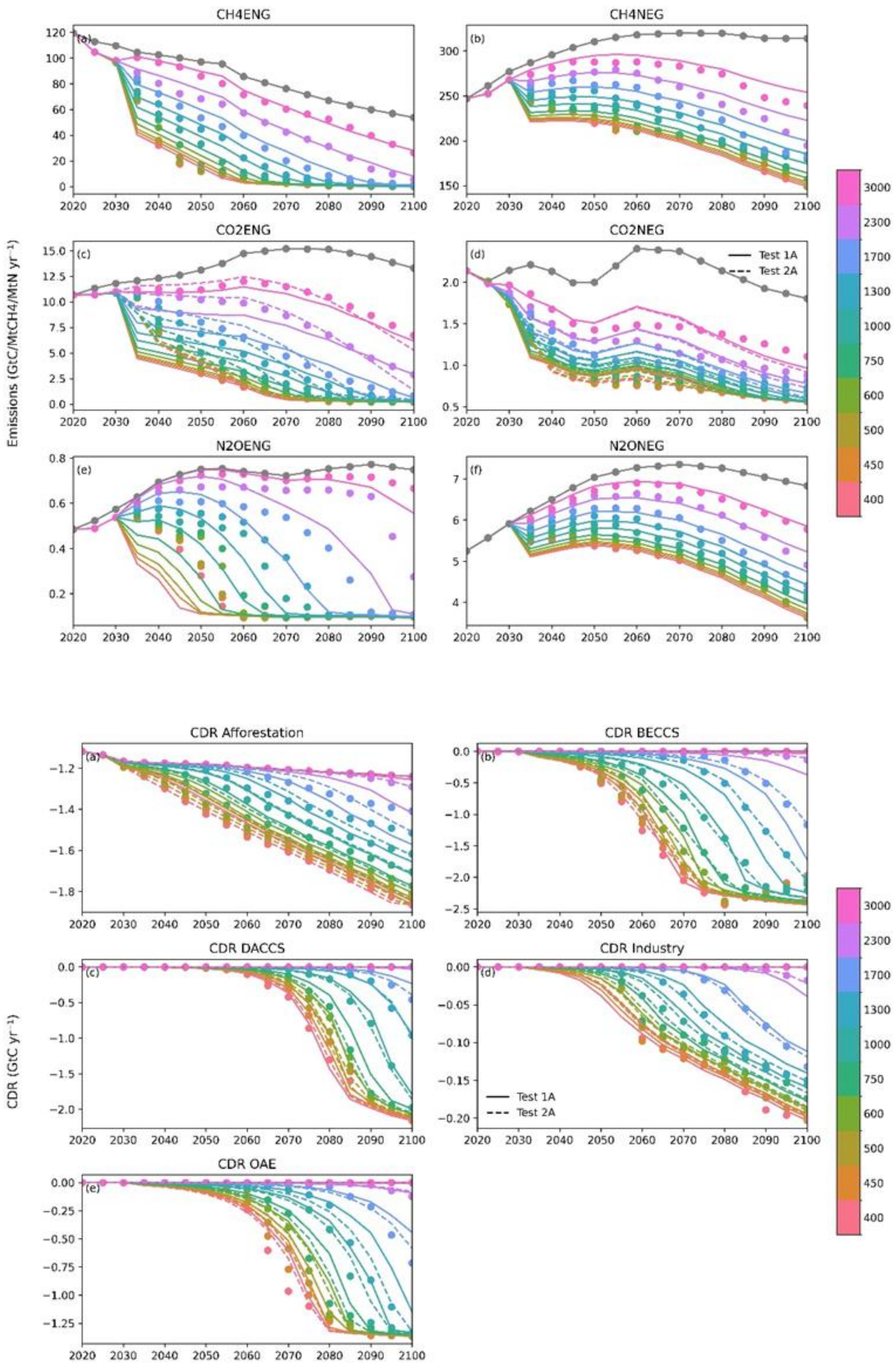

**Fig. 5 | Evaluating emIAM using the carbon price pathway (Test 1A, solid lines) and the cumulative carbon budget (Test 2A, dashed lines) as the target.** See Table 2 and the text for details of the tests.

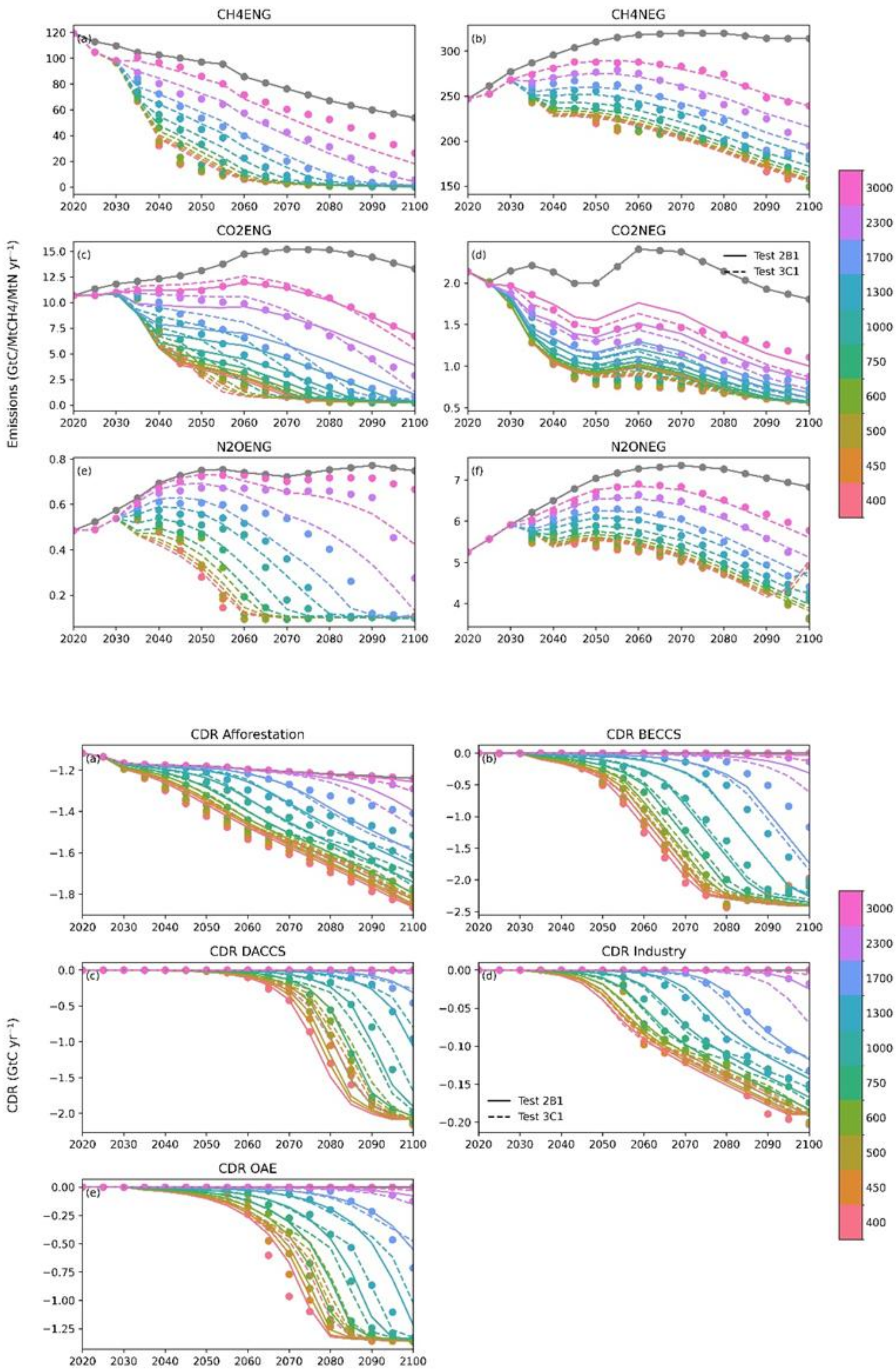

**Fig. 6 | Evaluating emIAM using the cumulative carbon budget (Test 2B1, solid lines) and the cumulative GHG budget (Test 3C1, dashed lines) as the target.** See Table 2 and the text for details of the tests.

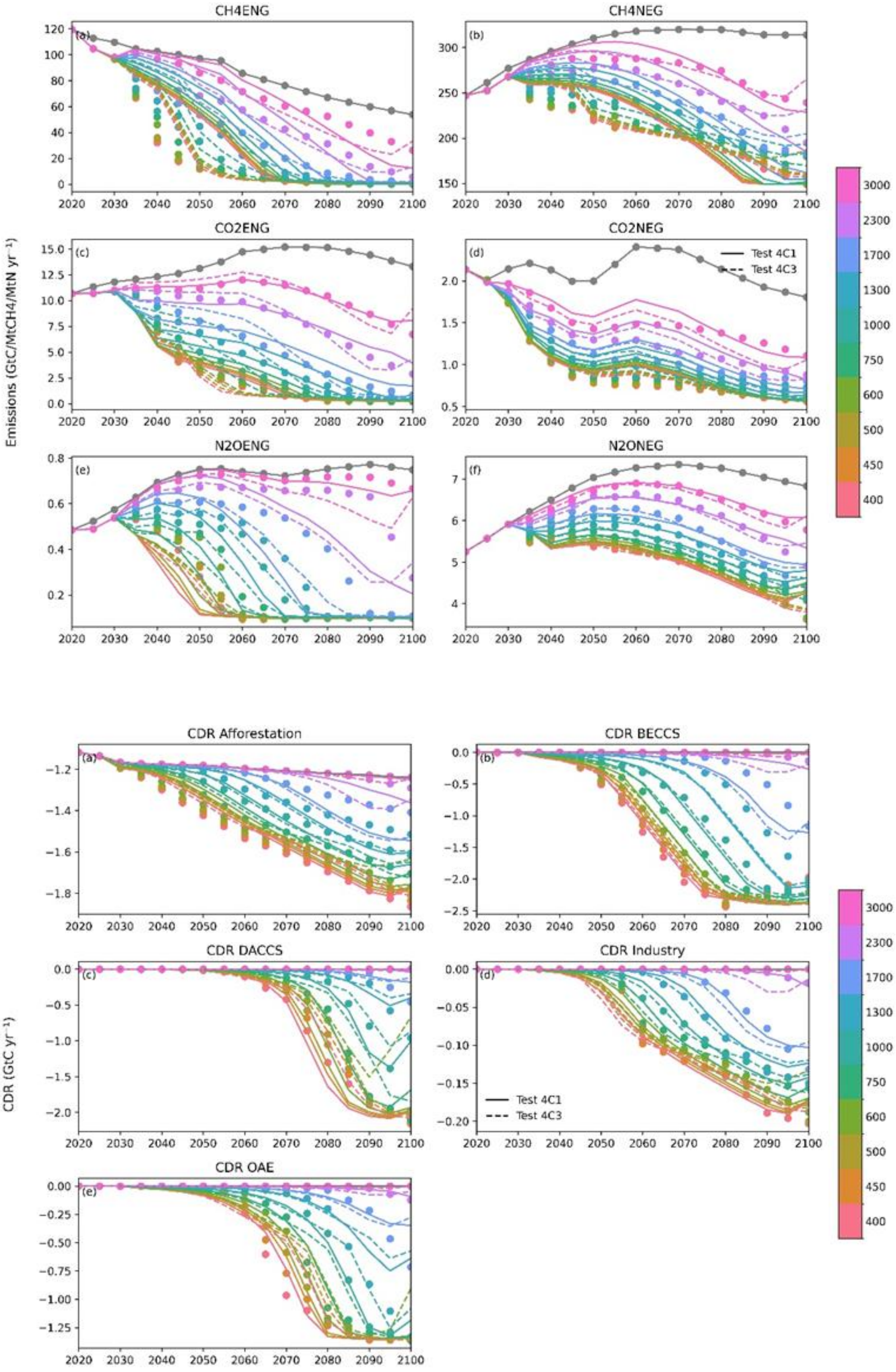

**Fig. 7 | Evaluating emIAM-ACC2 using the 2100 temperature as the target with uniform carbon price covering energy-related and non-energy-related emissions within each gas (Test 4C1, solid lines) and all emissions and removals across gases (Test 4C3, dashed lines).** See Table 2 and the text for details of the tests.

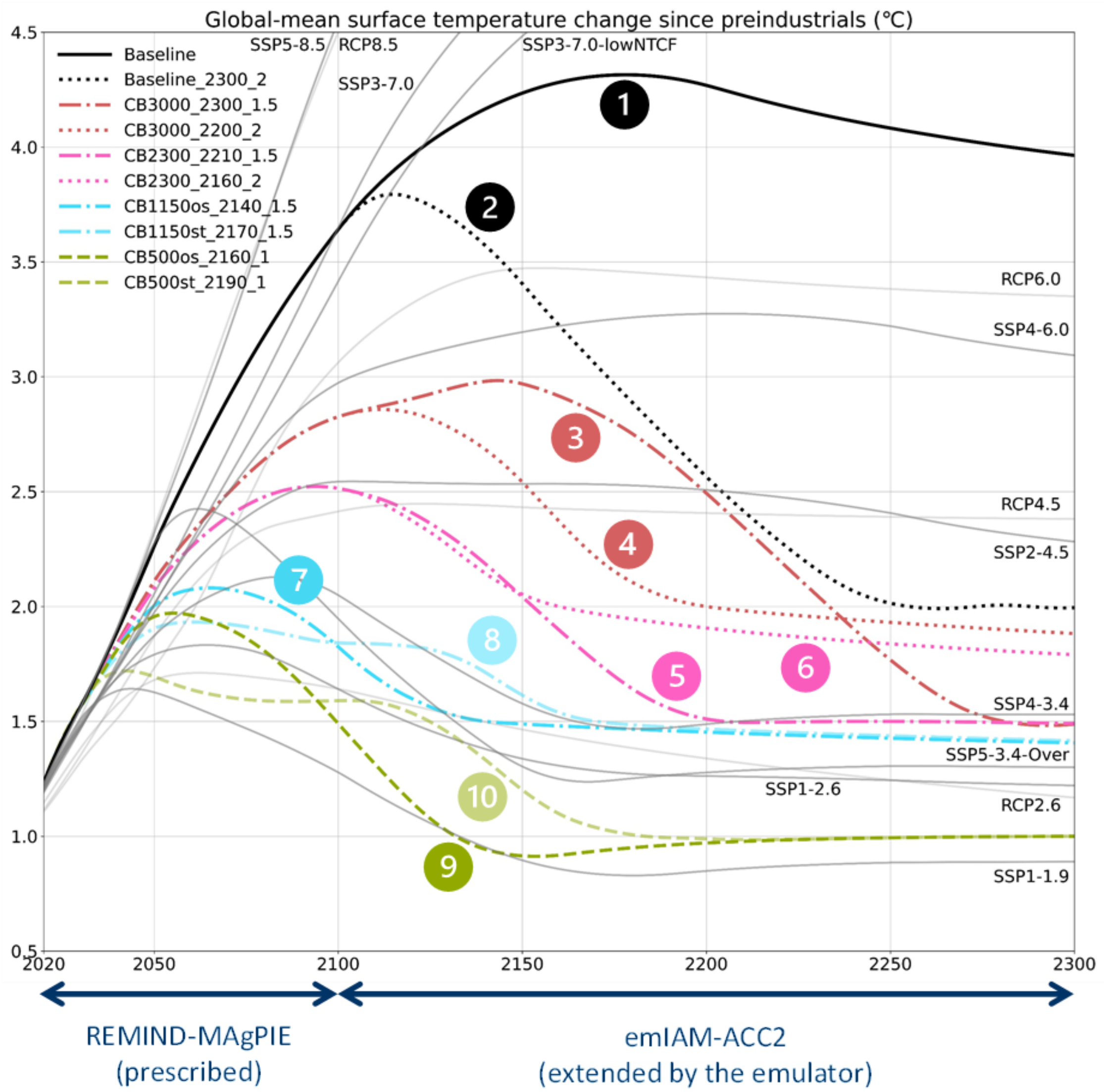

| ID | Pre-2100 scenario | Target level | Target year | Overshoot height | Overshoot period |
|---|---|---|---|---|---|
| 1 | Baseline | - | - | - | - |
| 2 | Baseline | 2 °C | 2300 | 1.79 °C | 213 year |
| 3 | CB3000 | 1.5 °C | 2300 | 1.48 °C | 253 year |
| 4 | CB3000 | 2 °C | 2200 | 0.86 °C | 154 year |
| 5 | CB2300 | 1.5 °C | 2210 | 1.02 °C | 177 year |
| 6 | CB2300 | 2 °C | 2160 | 0.52 °C | 113 year |
| 7 | CB1150os | 1.5 °C | 2140 | 0.58 °C | 112 year |
| 8 | CB1150st | 1.5 °C | 2170 | 0.43 °C | 142 year |
| 9 | CB500os | 1 °C | 2160 | 0.97 °C | 111 year |
| 10 | CB500st | 1 °C | 2190 | 0.72 °C | 163 year |

**Fig. 8 | Overview of the extended baseline and mitigation scenarios.** See text for details.

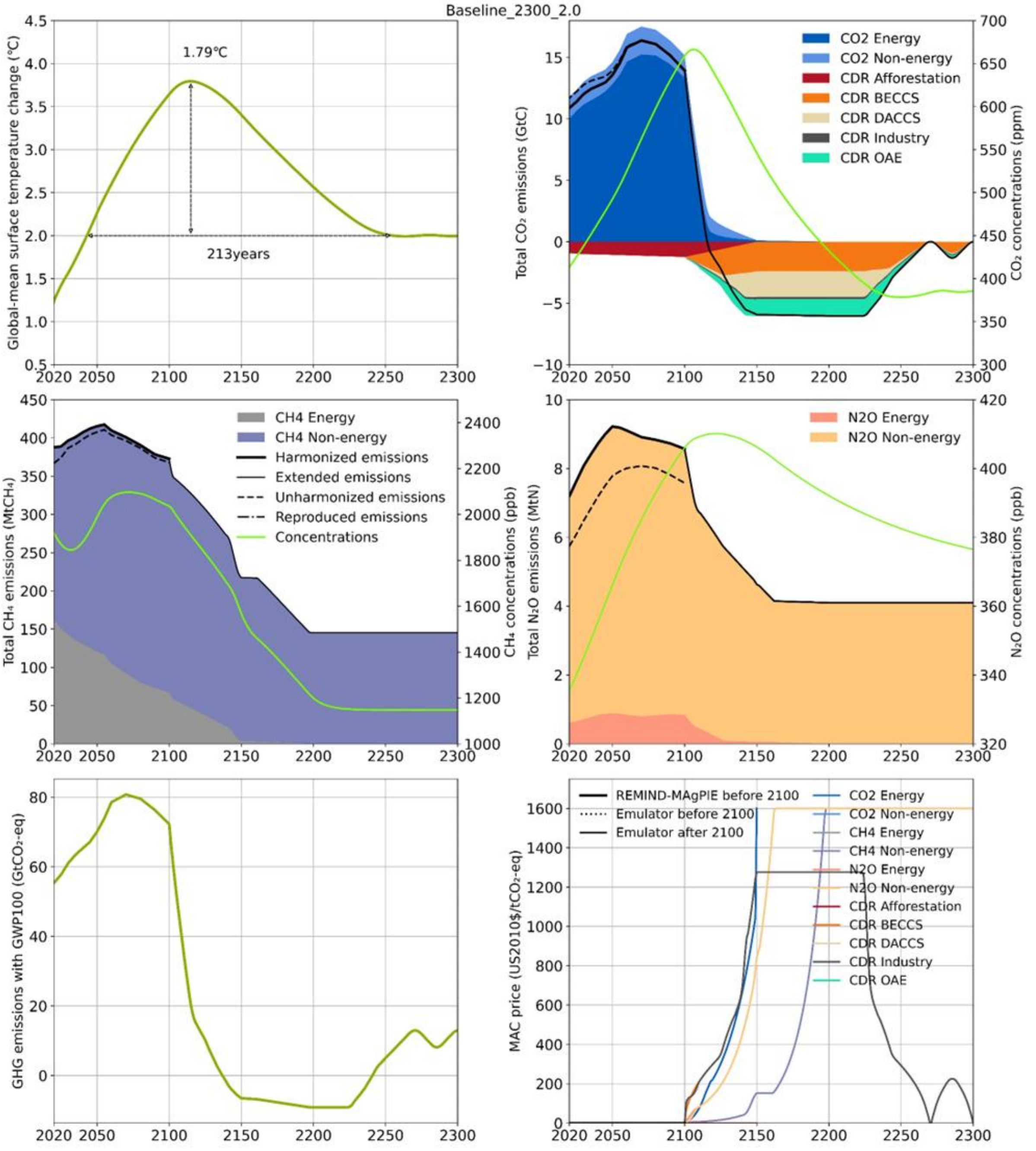

**Fig. 9 | Overview of Scenario #2: 2 °C scenario with overshoot till 2300 (peak warming of about 4 °C).**

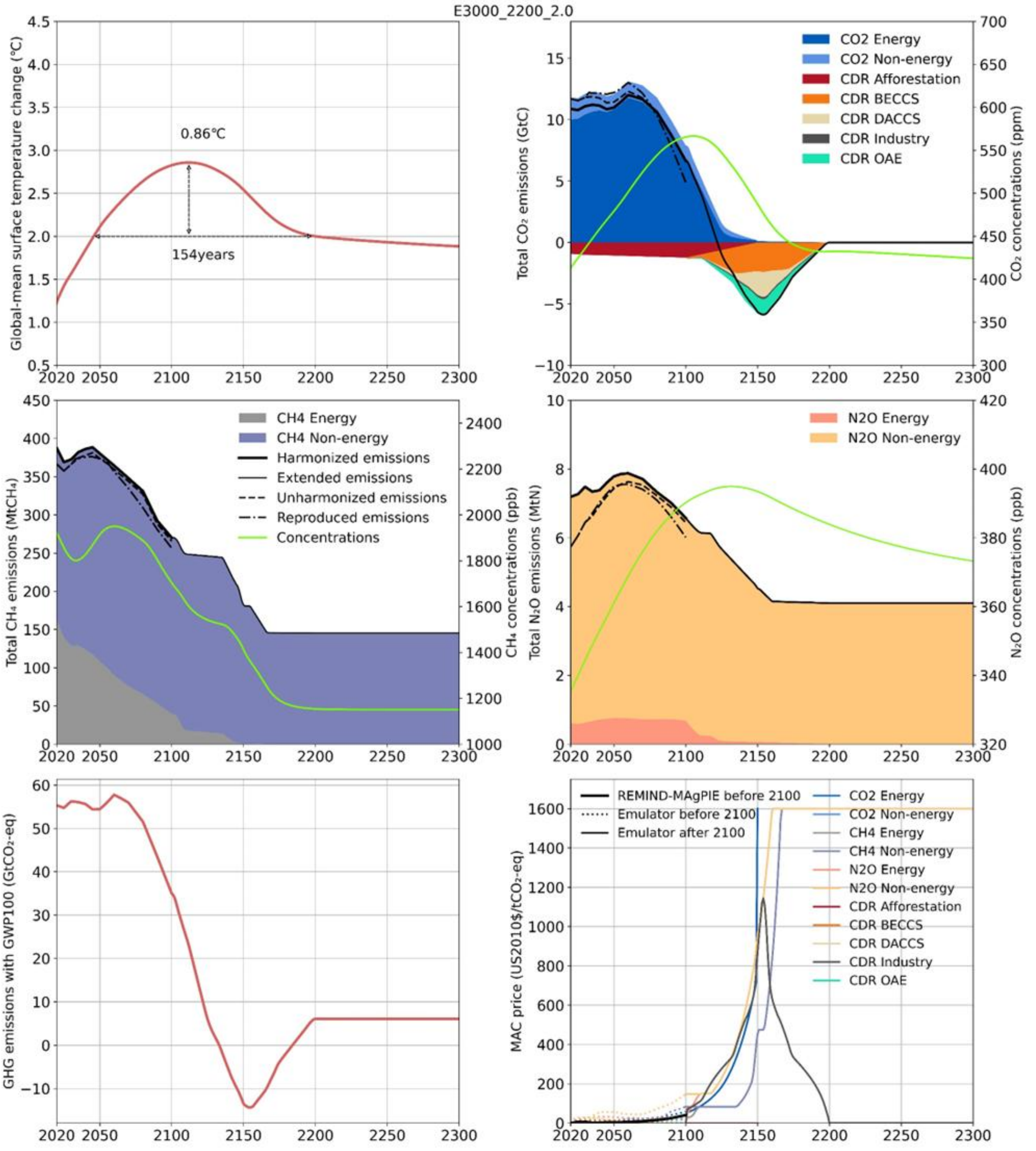

**Fig. 10 | Overview of Scenario #4: 2 °C scenario with overshoot till 2200 (peak warming of about 3 °C)**

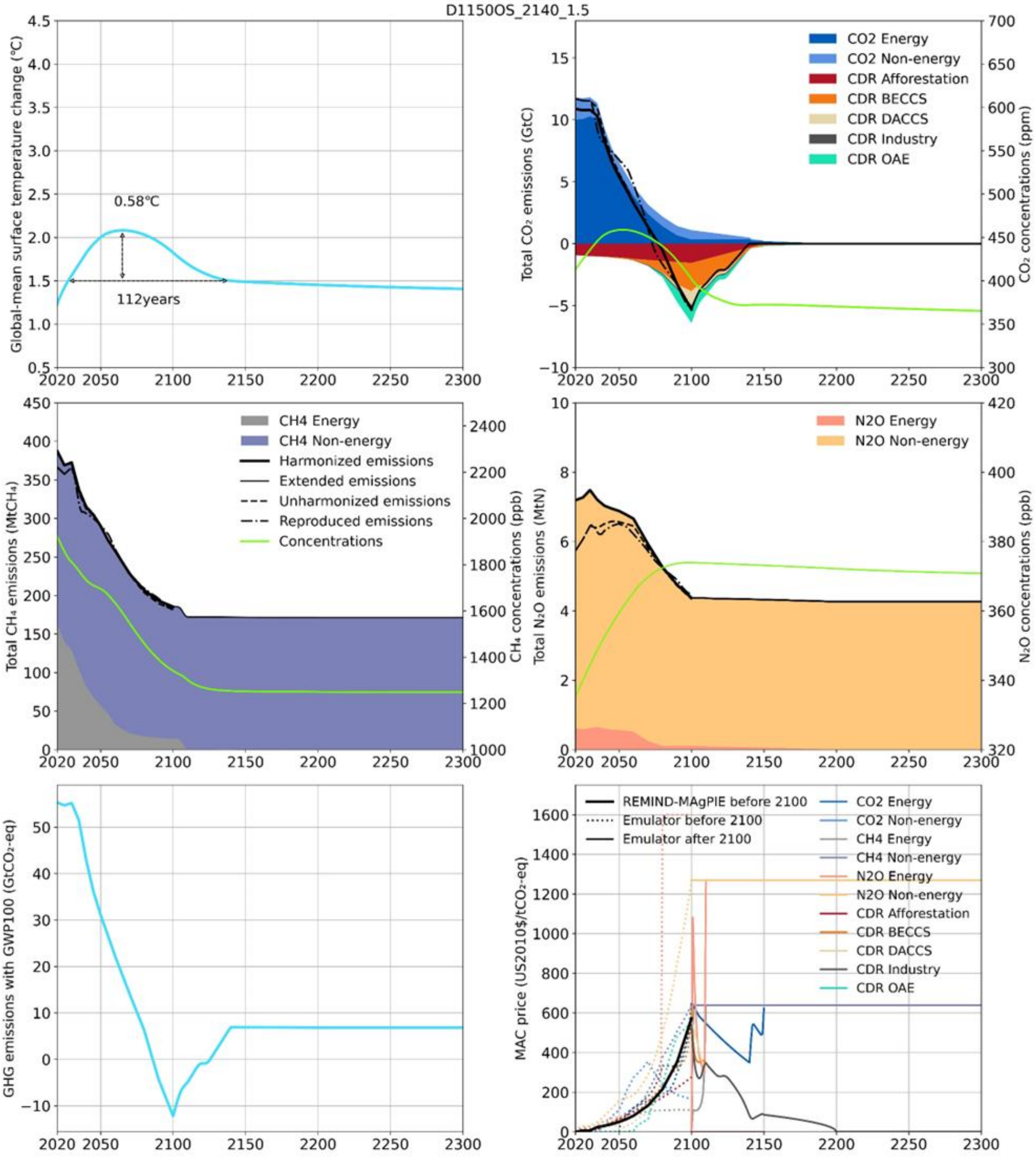

**Fig. 11 | Overview of Scenario #7: 1.5 °C scenario with overshoot till 2140 (peak warming of about 2 °C)**

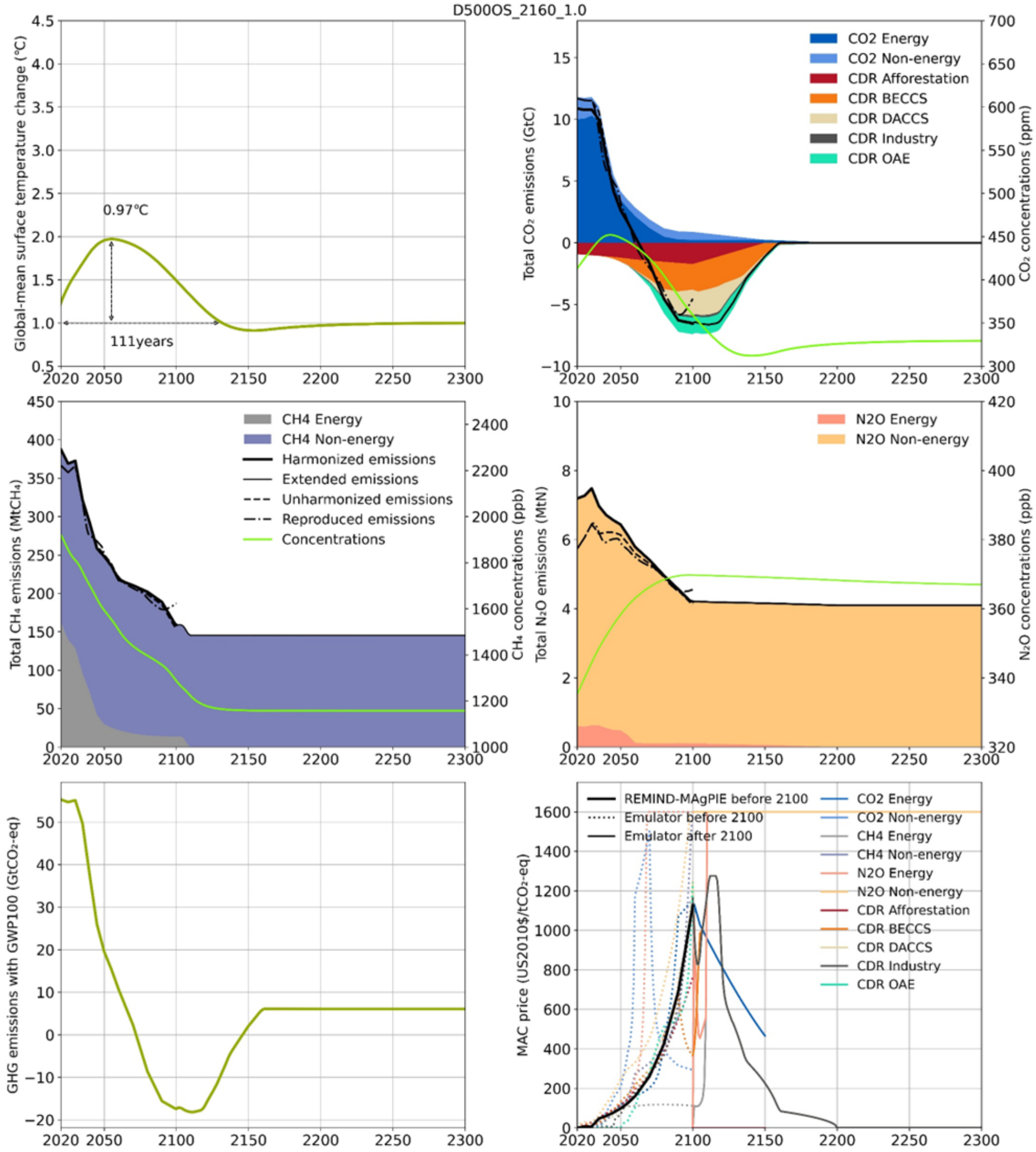

**Fig. 12 | Overview of Scenario #9: 1 °C scenario with overshoot till 2160 (about 1.5 °C warming in 2100)**

## Supplementary Information

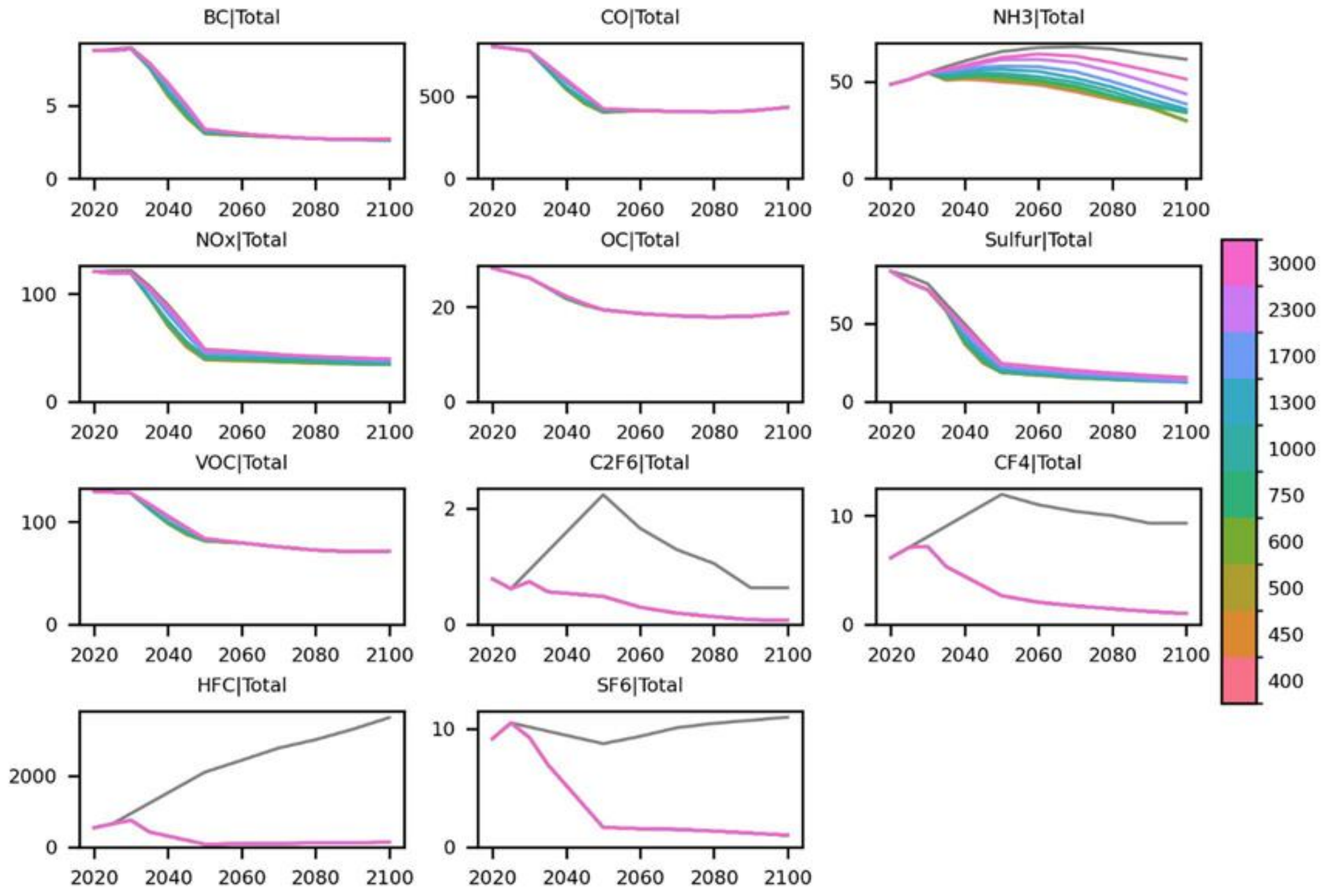

**Supplementary Fig. 1 | Sensitivity scenarios used for developing the REMIND-MAgPIE emulator.** The figure above shows the sensitivity scenarios of all climate forcers, excluding the three major GHGs considered in the emulator. Those of the three major GHGs are shown in Fig. 2.

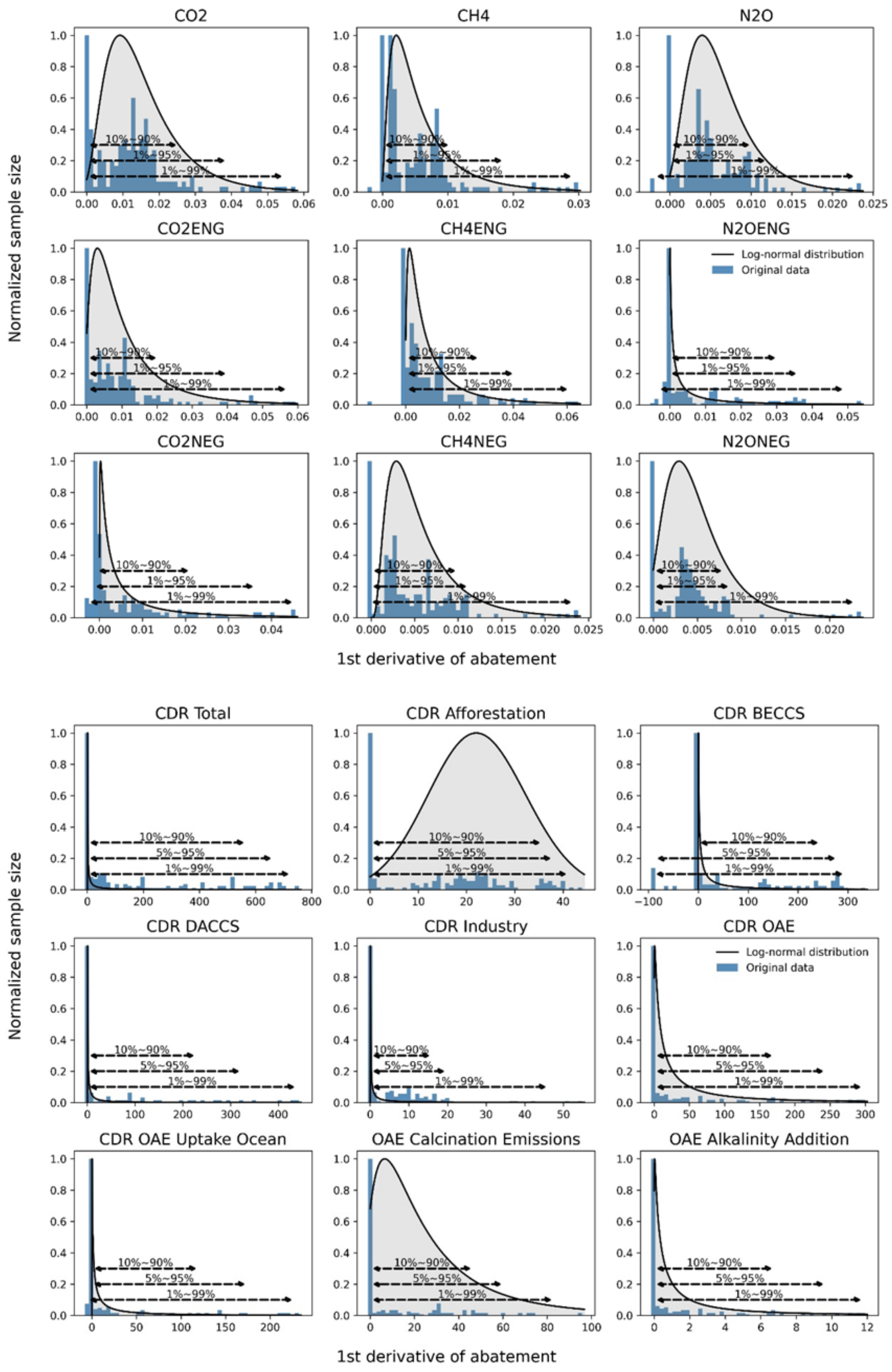

**Supplementary Fig. 2 | The first derivatives of temporal changes in abatement/removal levels indicated by the REMIND-MAgPIE sensitivity scenarios.**

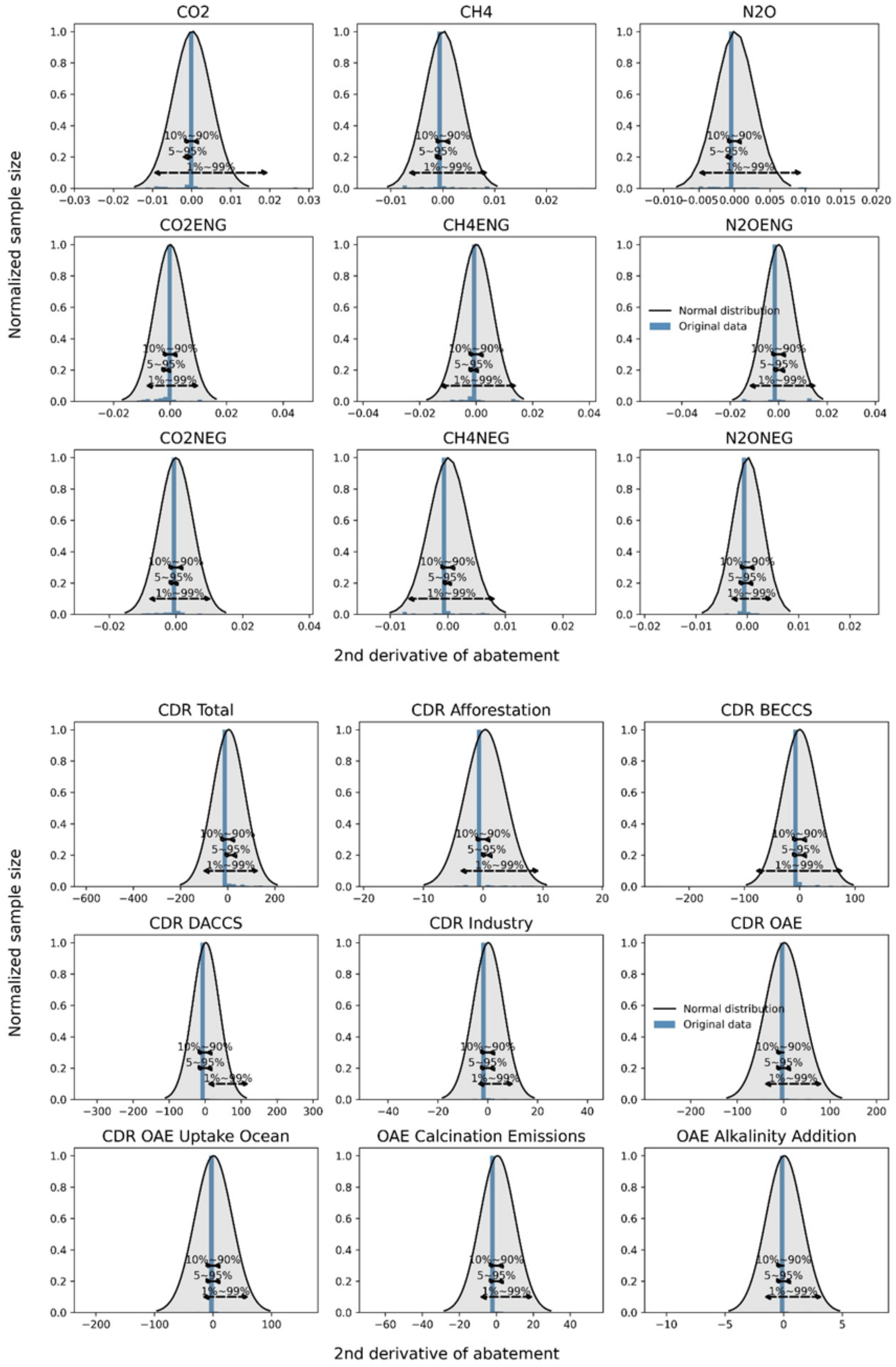

**Supplementary Fig. 3 | The second derivatives of temporal changes in abatement/removal levels indicated by the REMIND-MAgPIE sensitivity scenarios.**

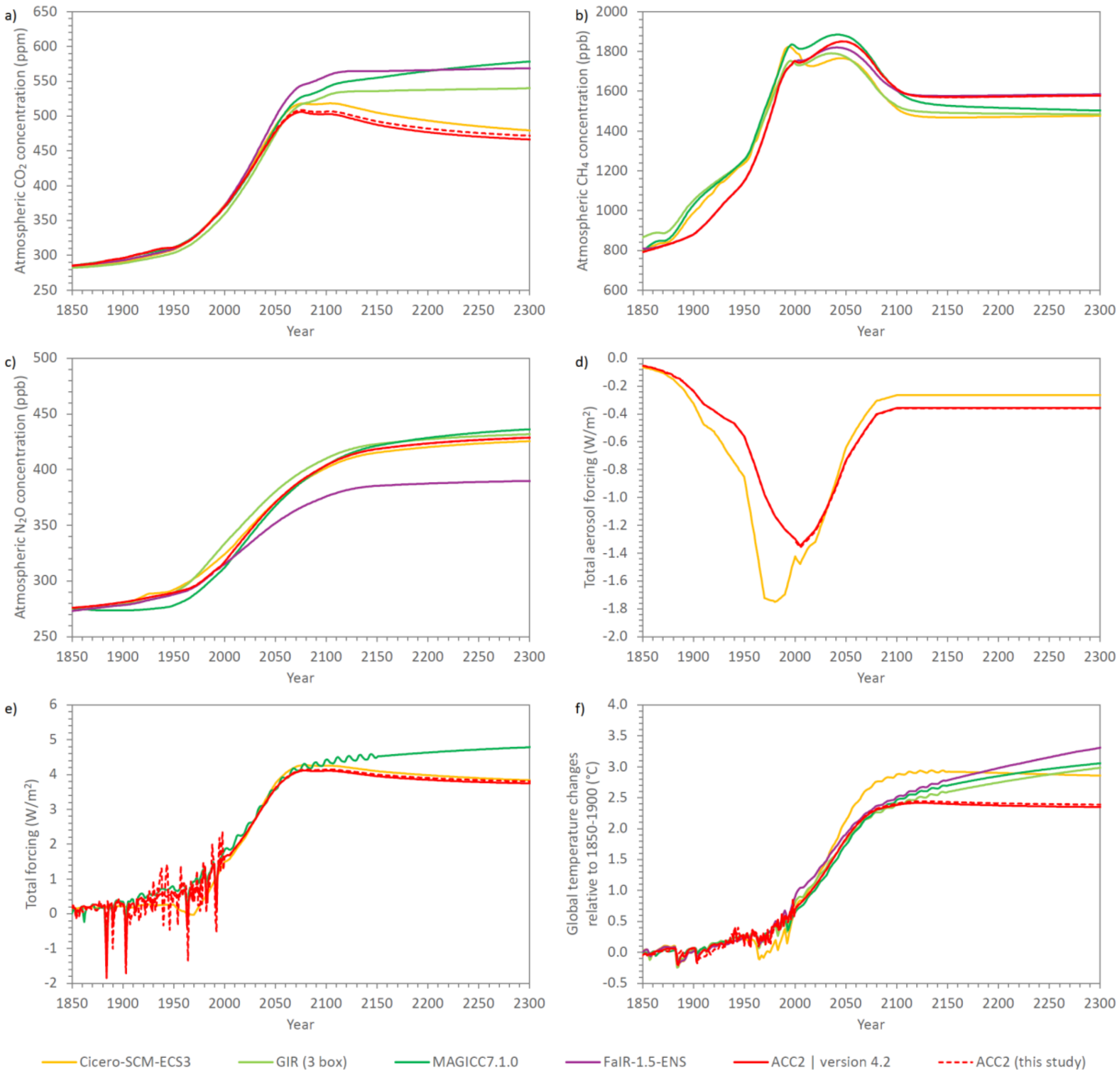

**Supplementary Fig. 4 | Model response of ACC2 in comparison with other reduced-complexity climate models simulating the RCP4.5 scenario.** See the caption of Fig. 4.

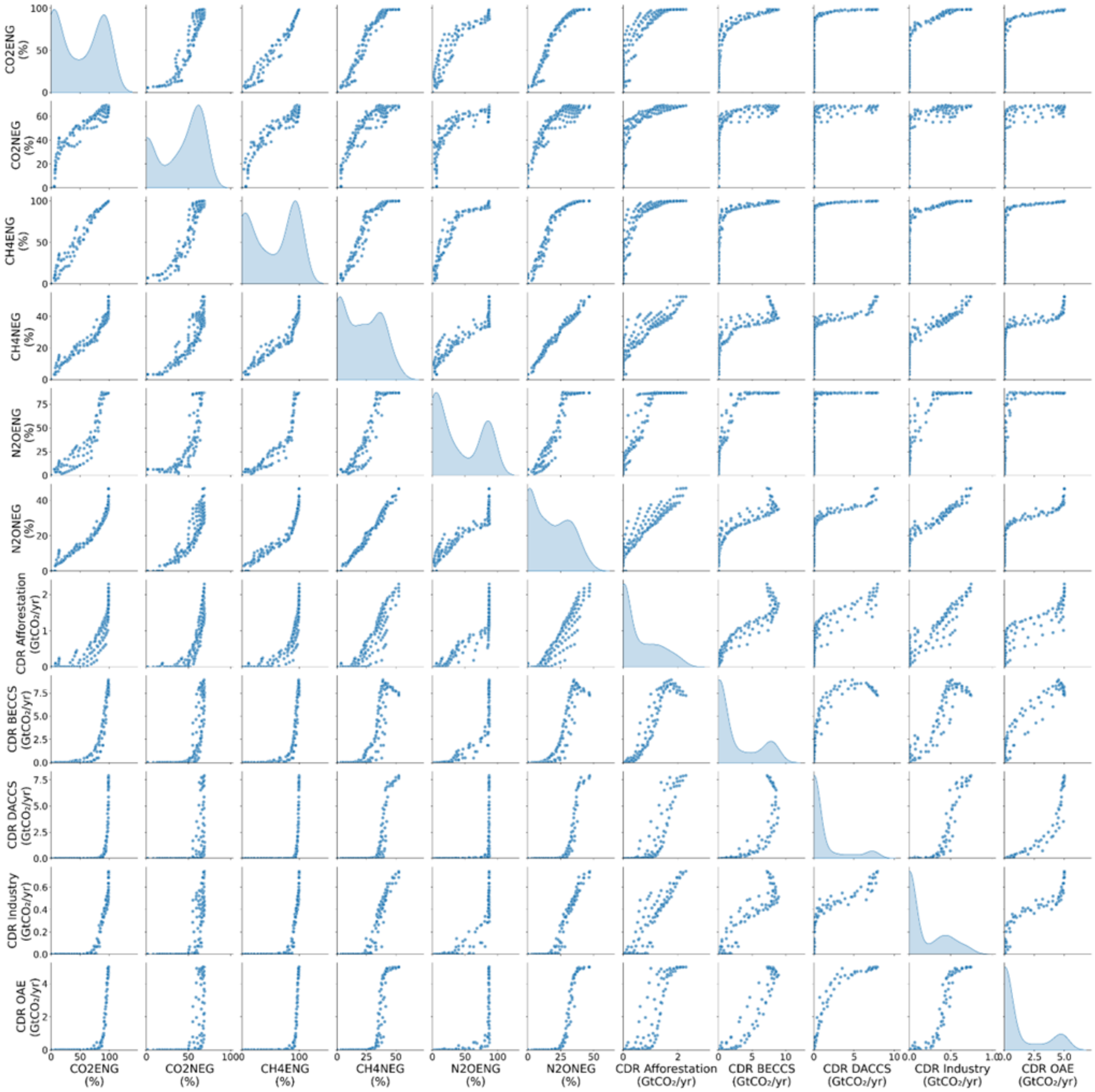

**Supplementary Fig. 5 | Correlations between different sectoral abatement and removal levels exhibited in the REMIND-MAgPIE sensitivity scenarios.**

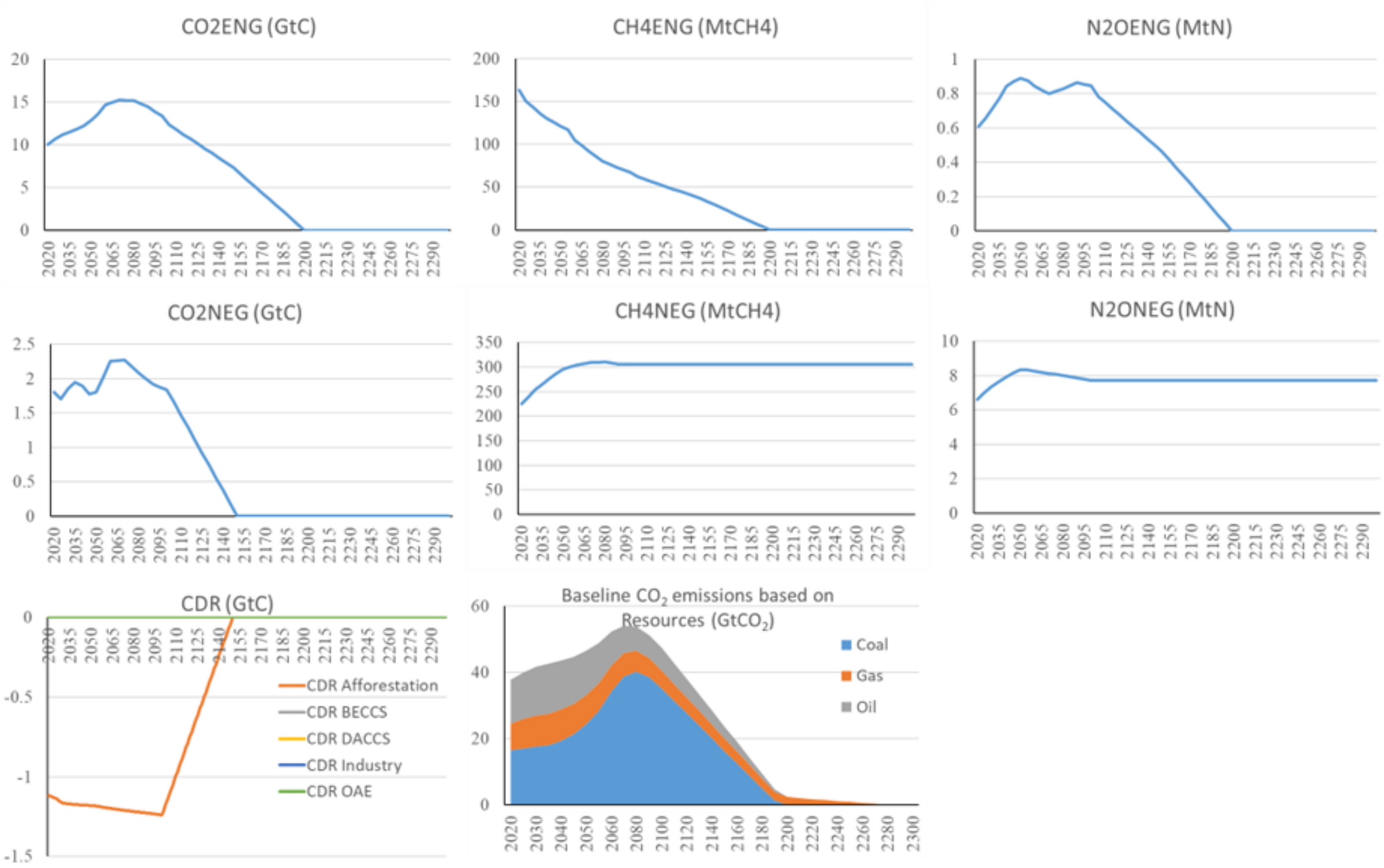

**Supplementary Fig. 6 | Extended baseline scenario by gas, sector, and CDR**

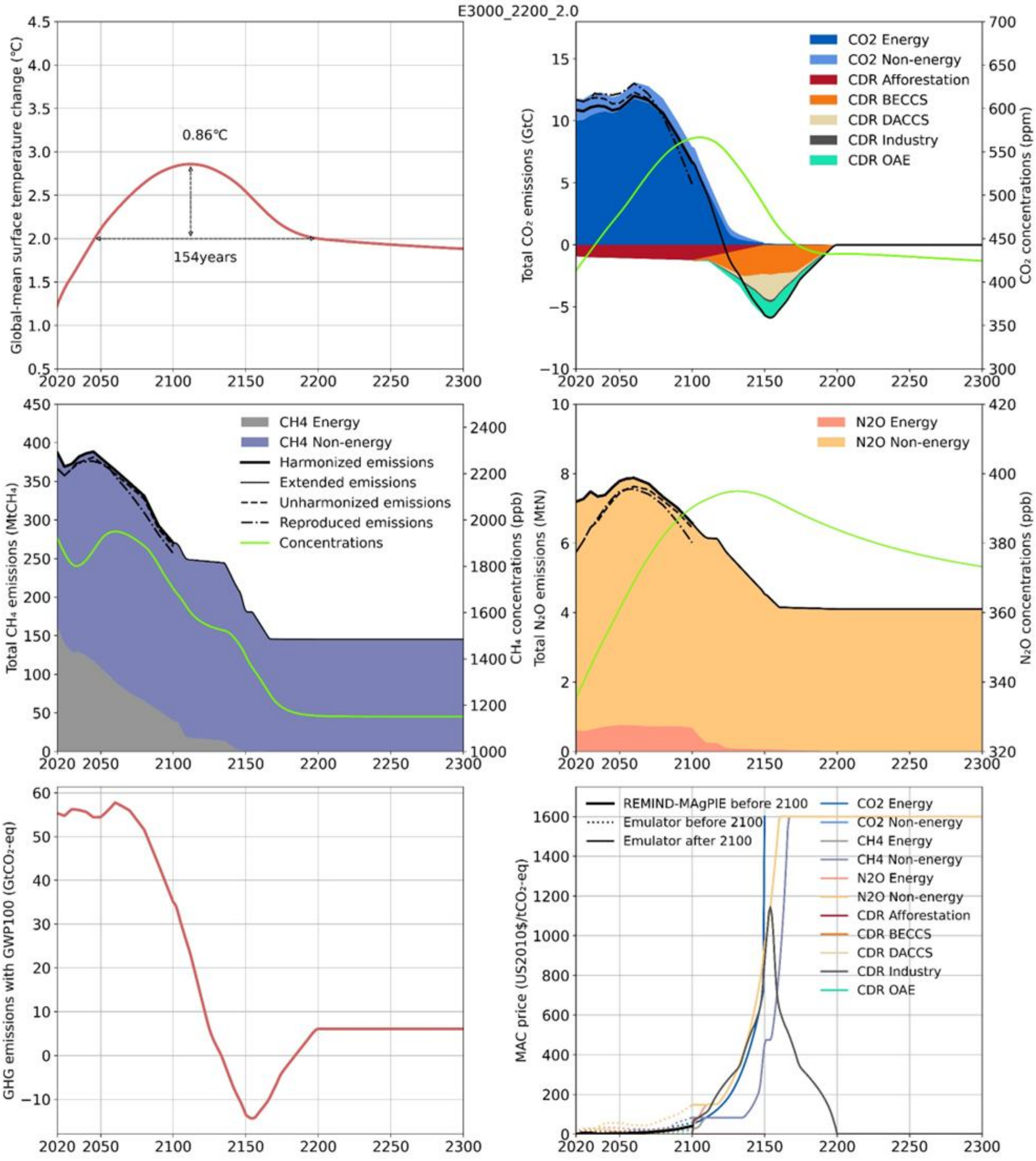

**Supplementary Fig. 7 | Overview of Scenario #3: 2 °C scenario with overshoot till 2200**

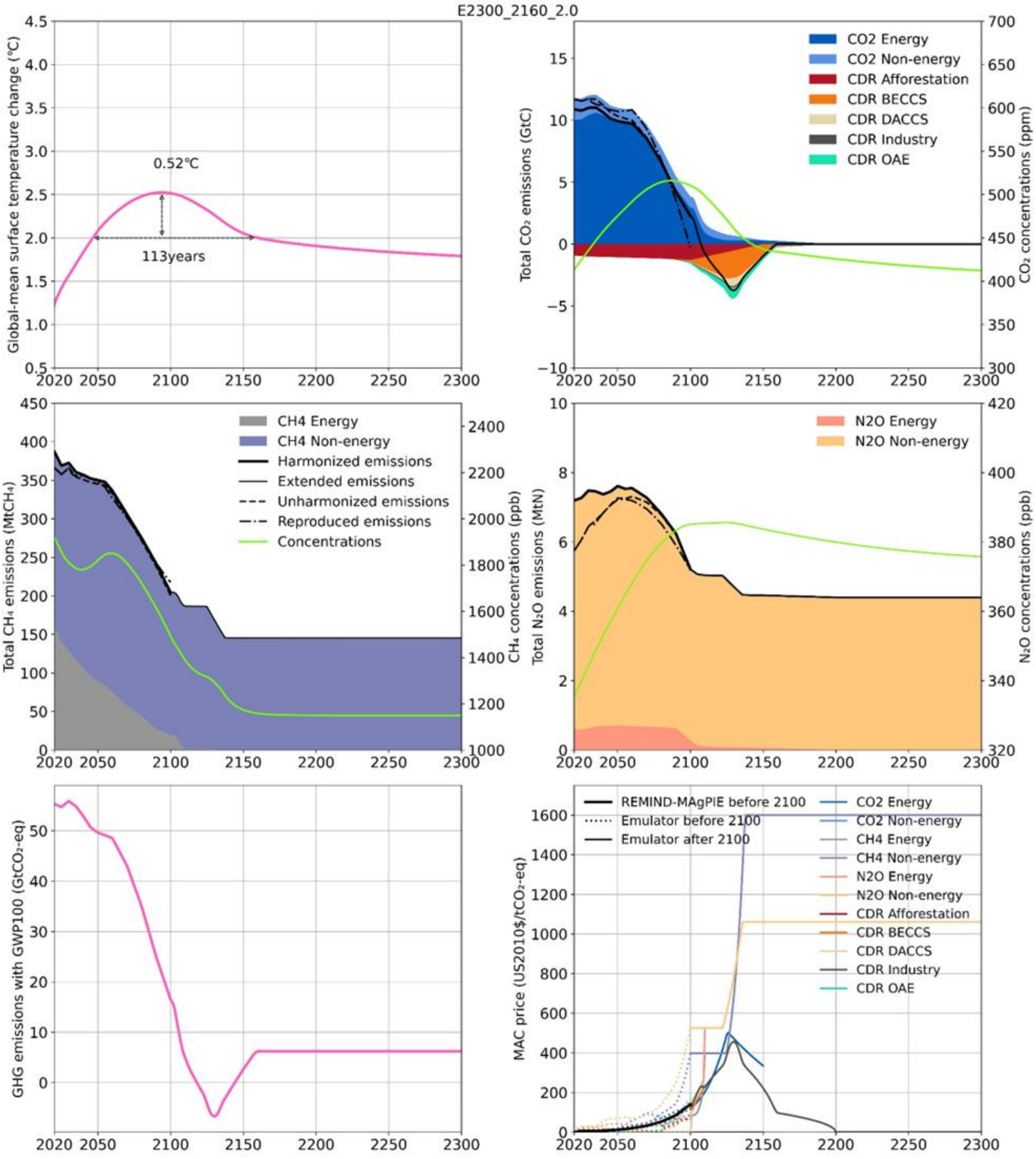

**Supplementary Fig. 8 | Overview of Scenario #5: 2 °C scenario with overshoot till 2160 (about 1.5 °C warming in 2100)**

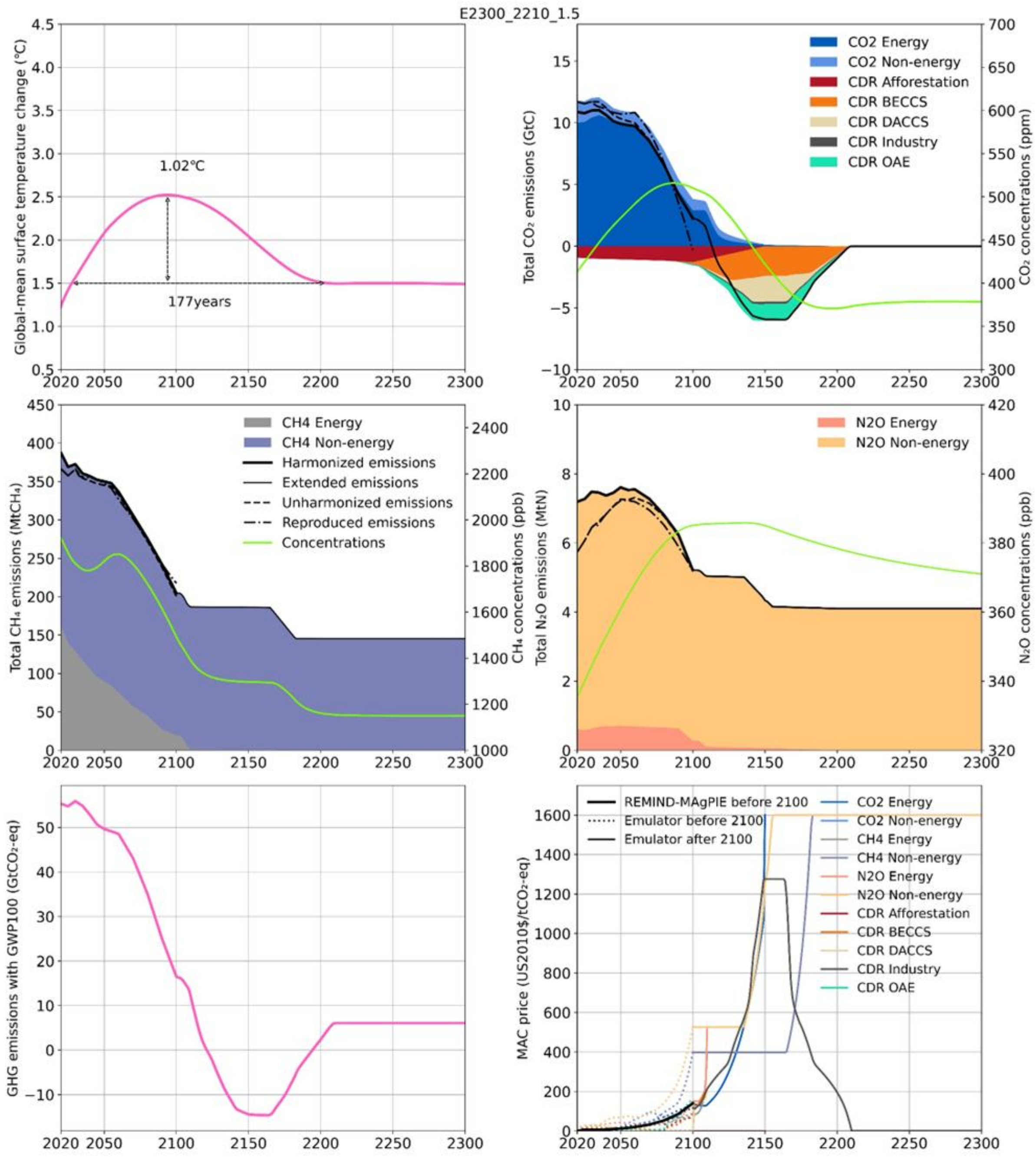

**Supplementary Fig. 9 | Overview of Scenario #6: 1.5 °C scenario with overshoot till 2210 (about 1.5 °C warming in 2100)**

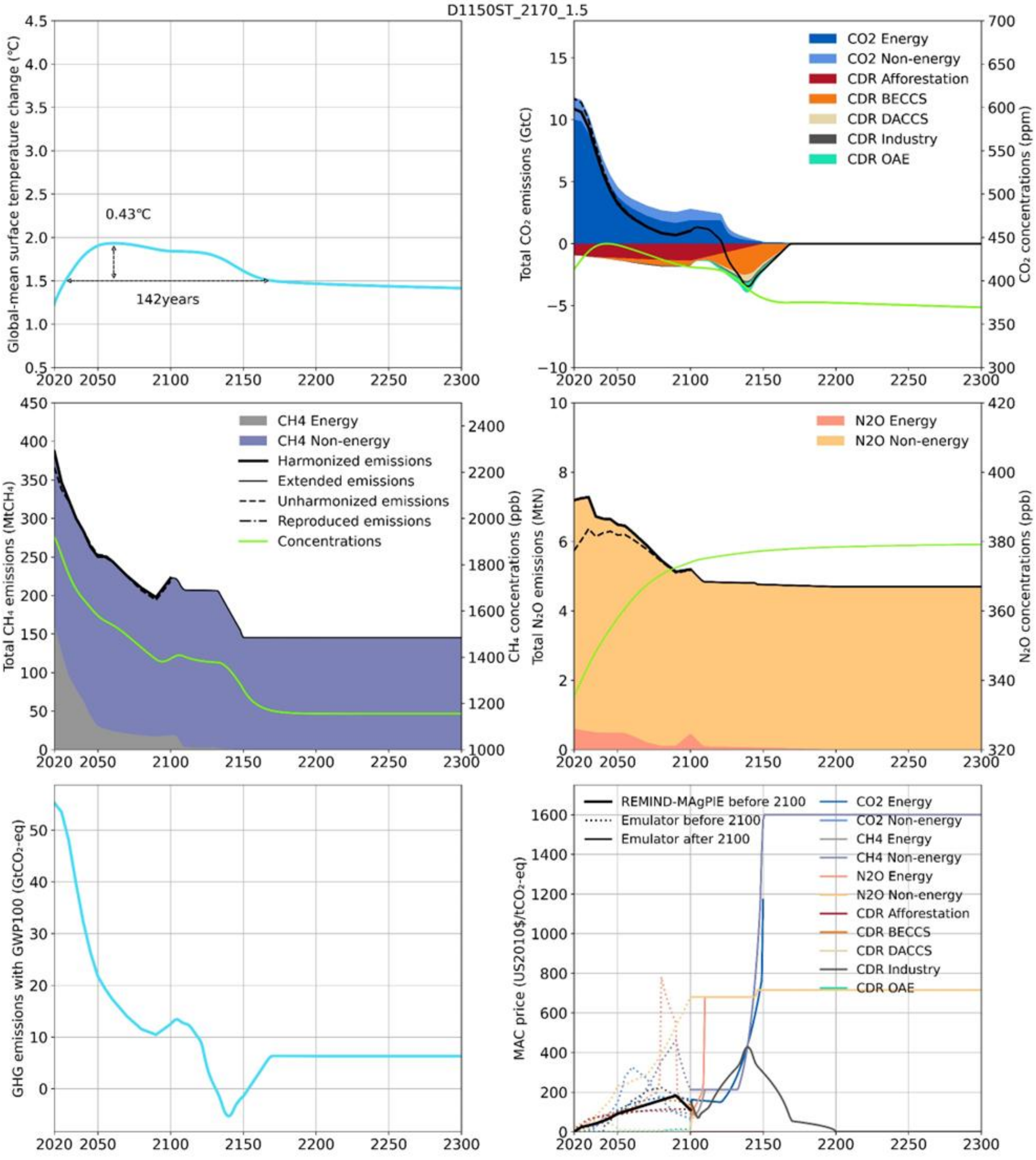

**Supplementary Fig. 10 | Overview of Scenario #8: 1.5 °C scenario with overshoot till 2170 (about 1.5 °C warming in 2100)**

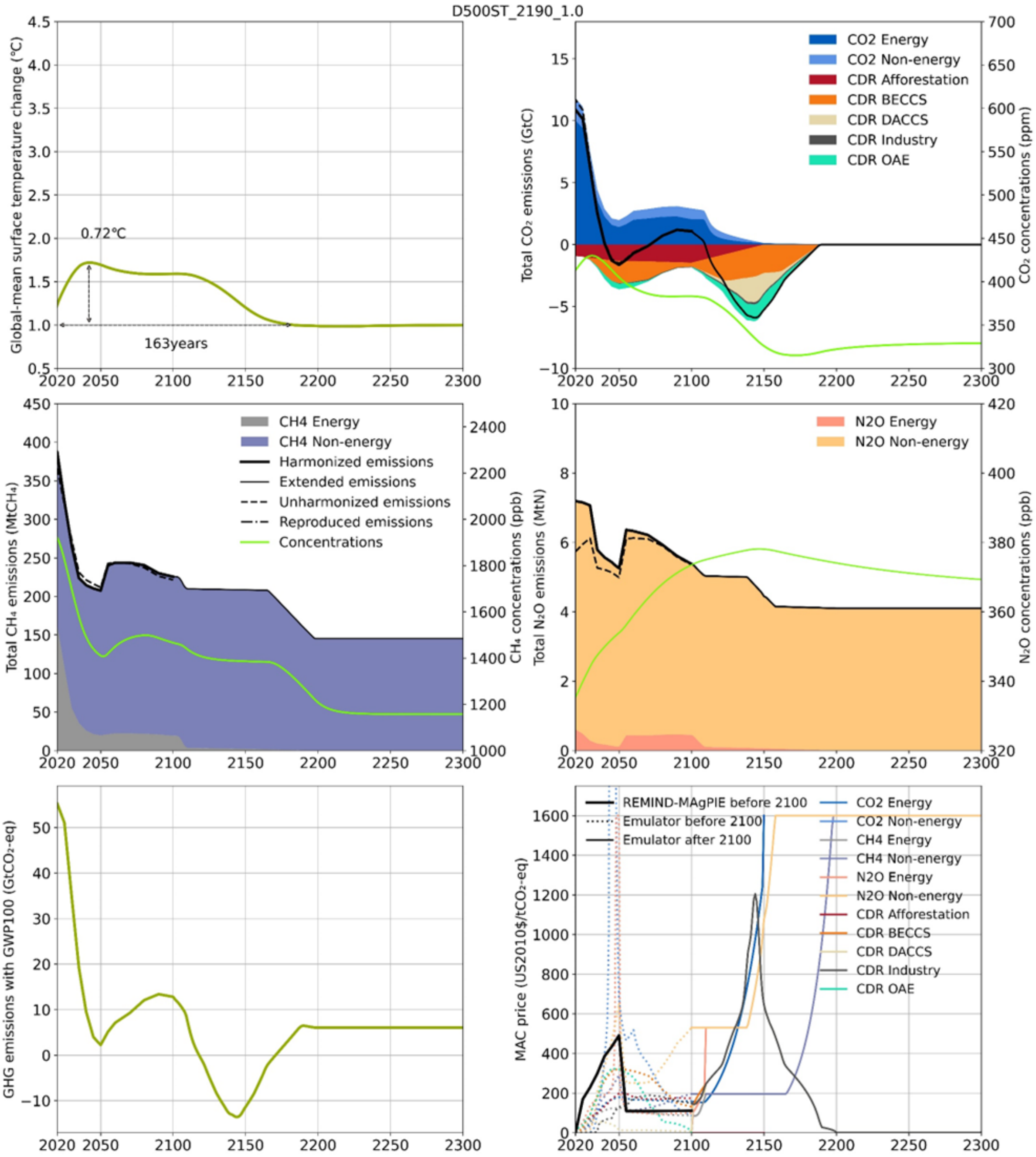

**Supplementary Fig. 11 | Overview of Scenario #10: 1 °C scenario with overshoot till 2190 (about 1.5 °C warming in 2100)**